\documentclass[AMS,STIX2COL,SYSTEMFONTS]{WileyNJD-v2}

\usepackage{amsmath,bm,bbm}
\usepackage{url} 
\usepackage{natbib}

\usepackage{graphicx}
\usepackage{float}
\usepackage{subcaption}
\usepackage{algorithm}
\usepackage{algpseudocode}
\usepackage{graphicx,psfrag,epsf}
\usepackage{enumerate}
\usepackage{dsfont}
\newcommand\omicron{o}
\usepackage{geometry}
\usepackage{marginnote}
\usepackage{xcolor}
\usepackage{comment}
\usepackage[colorinlistoftodos]{todonotes}

\newcommand{\N}{\mathrm{N}}

\articletype{Article Type}%

\received{26 April 2016}
\revised{6 June 2016}
\accepted{6 June 2016}

\begin{document}

\title{Regression-Based Bayesian Estimation and Structure Learning for Nonparanormal Graphical Models}

\author[1]{Jami J. Mulgrave*}

\author[2]{Subhashis Ghosal}

\authormark{Mulgrave, J. J. \textsc{et al}}

\address{\orgdiv{Department of Statistics}, \orgname{North Carolina State University}, \orgaddress{\state{North Carolina}, \country{USA}}}

\corres{*Jami Mulgrave, \email{jnj2102@gmail.com}}


\abstract[Summary]{A nonparanormal graphical model is a semiparametric generalization of a Gaussian graphical model for continuous variables in which it is assumed that the variables follow a
Gaussian graphical model only after some unknown smooth monotone transformations.  We consider a Bayesian approach to inference in a nonparanormal graphical model in which we put priors on the unknown transformations through a random series based on B-splines.  We use a regression formulation to construct the likelihood through the Cholesky decomposition on the underlying precision matrix of the transformed variables and put shrinkage priors on the regression coefficients. We apply a plug-in variational Bayesian algorithm for learning the sparse precision matrix and compare the performance to a posterior Gibbs sampling scheme in a simulation study.  We finally apply the proposed methods to a real data set.}


\maketitle


\section{Introduction}\label{Intro}

The Gaussian graphical model (GGM) is a mathematical model commonly used to describe conditional independence relationships among normally distributed random variables.  The estimation of the underlying graph in a GGM is known as structure learning.  Zeros in the inverse covariance matrix, or the precision matrix, indicate that the corresponding variables in the data set are conditionally independent given the rest of the variables in the data set, and this relationship is represented by the absence of an edge in the graph.  Similarly, nonzero entries in the precision matrix are represented by edges in the graph and correspond to conditionally dependent variables in the data set.  Thus, an assumed sparsity condition is used to learn the conditional dependence structure in a GGM. An extension of the GGM is the nonparanormal graphical model \citep{liu_nonparanormal:_2009} in which the random variables are replaced by transformed variables that are assumed to be normally distributed.  \cite{liu_nonparanormal:_2009} use a truncated empirical distribution function to estimate the functions and then estimate the precision matrix of the transformed variables using the graphical lasso.  
The Bayesian method for the nonparanormal graphical model \citep{mulgrave_bayesian_2020} uses a random series B-splines prior to estimate the functions and a Student-t spike-and-slab prior to estimate the resulting precision matrix. These extensions differ from the Gaussian copula graphical model \citep{pitt_efficient_2006,dobra_copula_2011,liu_high-dimensional_2012,mohammadi_bayesian_2017} in that the nonparanormal graphical model concurrently estimates the transformation functions and the precision matrices. Nonparanormal graphical model approaches have been applied to discrete data models of interactions between genes \cite{nguyen_structure_2018} and to test differential gene networks \cite{zhang_testing_2020}.

Estimation of a sparse precision matrix is necessary to learn the structure in GGMs and nonparanormal graphical models.  For unstructured precision matrices, a commonly used algorithm in the frequentist literature is the graphical lasso \citep{friedman_sparse_2008}.  A great number of algorithms have been proposed to solve this problem including \cite{meinshausen_high-dimensional_2006,yuan_model_2007,friedman_sparse_2008, banerjee_model_2008, daspremont_first-order_2008,rothman_sparse_2008, lu_smooth_2009, scheinberg_sparse_2010, witten_new_2011,mazumder_graphical_2012}. 

Analogous methods in the Bayesian literature use priors to aid the edge selection procedure. For instance, off-diagonal entries of the precision matrix may be set to zero by allowing a point mass at zero in the prior \citep{banerjee_bayesian_2015}, but the posterior is harder to compute or sample from. A normal spike-and-slab prior \citep{wang_scaling_2015} replaces the point mass at zero by a highly concentrated normal distribution around zero and similarly, a Laplace spike-and-slab prior \citep{gan_bayesian_2019} has been used. From a computational point of view, continuous shrinkage priors such as the horseshoe prior \citep{carvalho_handling_2009}, the Dirichlet-Laplace prior \citep{bhattacharya_dirichlet-laplace_2015}, and generalized double exponential prior \citep{armagan_generalized_2013}, bring in the effects of both a point mass and a thick tail by a single continuous distribution with an infinite spike at zero.

Ideally, we seek solutions that guarantee a sparse positive definite matrix using continuous shrinkage priors.  Since continuous shrinkage priors do not assign exact zeros, a variable selection procedure needs to be used to determine which of the small and nonzero elements should be specified as exactly zero.  Methods that use spike and slab priors naturally incorporate variable selection, whereas methods that use alternative priors need a thresholding procedure.  However, post-hoc thresholding procedures do not guarantee a positive definite precision matrix.  The methods in \cite{wang_scaling_2015} and \cite{wang_bayesian_2012} guarantee a positive definite matrix by way of the sampling algorithm.  \cite{wang_bayesian_2012} and \cite{peterson_inferring_2013} use the double exponential prior and improve on its use for sparsity by allowing each double exponential prior to have its own shrinkage parameter.   More recent methods estimate the inverse covariance matrix by using the normal spike and slab prior \citep{wang_scaling_2015,peterson_joint_2016,li_using_2019,li_expectation_2019} for variable selection in the graphical model context.  Lastly, \citep{williams_bayesian_2018} constructs Gaussian graphical models by estimating the partial correlation matrix using a horseshoe prior for regularization and for sparsity, using projection predictive selection, a method that allows for variable exclusion based on predictive utility, with good results. 

Utilizing a Cholesky decomposition is an alternative way to incorporate the positive definiteness constraint on precision matrices, but is very dependent on the ordering of the variables \citep{pourahmadi_covariance_2011}. We consider a prior based on Cholesky decomposition of the precision matrix that reduces this dependence. We derive a sparsity constraint that ensures a weak order invariance in that it maintains the same order of sparsity in the rows of the precision matrix by increasing the order of sparsity down the rows of the lower triangular matrix. We construct a pseudo-likelihood through regression of each variable on the preceding ones. The approach splits the very high dimensional original problem to several lower dimensional ones.   The method in \cite{wong_adaptive_2013} is also based on Cholesky decomposition, but it uses a noninformative Jeffreys' prior and the ordering issue of the Cholesky decomposition is not addressed.  

We consider two different priors, the horseshoe and the Bernoulli-Gaussian \citep{soussen_bernoulli-gaussian_2011}.   These priors have clear interpretations of the probability of nonzero elements \citep{soussen_bernoulli-gaussian_2011,van_der_pas_horseshoe_2014}, which allows us to effectively calibrate sparsity. The strength of the Bernoulli-Gaussian prior is that it leads to a sparse positive definite precision matrix that does not require thresholding and the strength of the horseshoe prior is that it is a better model of sparsity than the Bernoulli-Gaussian prior due to its heavier tails. Horseshoe priors have not yet been used for Bayesian nonparanormal graphical models that use transformation functions.  We compare the performance of the methods using both a variational Bayesian algorithm and a full Markov chain Monte Carlo (MCMC) sampling scheme.  Mean field variational Bayes \citep{jordan_introduction_1999,wainwright_graphical_2007} is an alternative to MCMC that allows for faster fitting by deterministic optimization.  A variational Bayesian method for Gaussian graphical models is developed in \cite{chen_bayesian_2011} and an expectation conditional-maximization approach is used by \cite{li_expectation_2019} in Gaussian copula graphical models. This approach has not yet been explored in the setting of a nonparanormal graphical model.   We wish to determine if we can retain the information learned in a Bayesian nonparanormal graphical model while speeding up the estimation process using variational Bayesian techniques.

The paper is organized as follows. In the next section, we describe the model and the sparsity constraint.  In Section \ref{VariationalBayesEstimation}, we describe the variational Bayesian algorithm.
In Sections \ref{MCMCHorseshoe} and \ref{MCMCBG}, we discuss particular priors and their corresponding Markov Chain Monte Carlo algorithms.  In Section \ref{Thresholding}, we describe a thresholding procedure and in Section \ref{ChoicePriorParams}, we detail the tuning procedure.  In Section \ref{SimulationResults}, we present a simulation study.  In Section \ref{RealData}, we describe a real data application. 

\section{Model and Priors}\label{ModelPriors} 
\subsection{Nonparanormal Transformation}\label{npn}
\begin{definition}
\label{def:nonparanormal}\rm 
A random vector $\bm{X} = (X_1,\ldots,X_p)^T$ has a nonparanormal distribution if there exist smooth monotone functions $\{f_{d}: d=1,\ldots,p\}$ such that $\bm{Y} = \bm{f}(\bm{X}) \sim \N_p(\boldsymbol\mu, \boldsymbol\Sigma)$, a normal distribution with mean $\boldsymbol\mu$, covariance matrix $\boldsymbol\Sigma$, and dimension $p$, and where $\bm{f}(\bm{X}) = (f_{1}(X_1),\ldots,f_{p}(X_p))'$. In this case we shall write $\bm{X}\sim \mathrm{NPN}(\bm{\mu},\bm{\Sigma},\bm{f})$. 
\end{definition}

We put prior distributions on the unknown transformation functions through a random series based on B-splines. In  \cite{mulgrave_bayesian_2020}, we have described the prior distributions, including the motivation and support for the choices made, in greater detail.  We briefly describe the prior in this section.  We represent the transformation functions 
$\bm{f}(\bm{x}) = (f_{1}(x_1),\ldots,f_{p}(x_p))'$ in a nonparanormal model $\bm{X}\sim \mathrm{NPN}(\bm{\mu},\bm{\Sigma},\bm{f})$ through a basis expansion 
\begin{equation}
\label{spline model}
f_d(x_d) = \sum_{j=1}^J\boldsymbol\theta_{dj}B_j(x_d) 
\end{equation}
where each $\theta_{dj}$ are coefficients, $B_j(\cdot)$ are the B-spline basis functions, $d = 1, \ldots, p$, $j = 1, \ldots J$,
and $J$ is the number of B-spline basis functions used in the expansion.  We assume that the precision matrix $\bm{\Omega} = \bm{\Sigma}^{-1}$ is sparse, in that, most of its off-diagonal entries are zero.  However, the model is not identifiable, since location-scale changes in the transformation functions and the normal distributions can be cancelled by each other. To resolve the issue, one possibility is to fix the mean-vector to zero and assume that the covariance matrix is a correlation matrix, but putting a prior on such a matrix maintaining sparsity of its inverse appears inconvenient.  Therefore, we let the mean and the precision matrix be free parameters while putting restrictions on the transformations.  We begin with a normal prior on each of the coefficients
of the B-splines, $\bm{\theta}_d = (\theta_{d1}, \ldots, \theta_{dJ})'$, that is set to be $\bm{\theta}_d \sim \N_J(\boldsymbol\zeta, \omicron^2 \bm{I})$, where $\omicron^2$ is some positive constant, $\boldsymbol\zeta$ is some vector of constants, and $\bm{I}$ is the identity matrix, and impose a monotonicity restriction on them to make the transformation $f_d$ monotone (see below for details).  We impose the following
two linear constraints on the coefficients through function values of the transformations:
$0 = f_d(1/2)=\sum_{j=1}^J\theta_{dj}B_j(1/2)$ and $1 = f_d(3/4)-f_d(1/4)=\sum_{j=1}^J\theta_{dj}[B_j(3/4)-B_j(1/4)].$  The linear constraints can be written in matrix/vector form as $\bm{A}\boldsymbol\theta_d = \bm{c}$ for each $d = 1, \ldots, p$.  The linear nature of the constraints allow us to retain the joint normality of the coefficient vectors before the monotonicity restriction, and hence a truncated joint normal after the restriction is imposed.

By the properties of a B-spline basis function, if the B-spline coefficients, $\theta_{dj}$ are increasing in $j$, then $f_j$ is an increasing function.  We thus impose the monotonicity constraint on the coefficients, which is equivalent with the series of inequalities $\theta_{d2}- \theta_{d1} > 0,\ldots, \theta_{dJ} - \theta_{d, J - 1} > 0$. The monotonicity constraint can be expressed in matrix/vector form as $\bm{F}\bm{\theta}_d > \mathbf{0}$ for each $d = 1, \ldots, p$.  Thus, the prior on the coefficients before the truncation is imposed is given by $ \bm{\theta}_d|\{\bm{A}\bm{\theta}_d = \bm{c}\} \sim \N_J(\boldsymbol\xi,\boldsymbol\Gamma),$
where the prior mean and variance are
$
\boldsymbol\xi = \boldsymbol\zeta + \bm{A}'(\bm{A}\bm{A}')^{-1}(\bm{c}-\bm{A} \boldsymbol\zeta) 
\label{conditional mean} $ and $
\boldsymbol\Gamma = \omicron^{2}[\bm{I}-\bm{A}'(\bm{A}\bm{A}')^{-1}\bm{A}]
\label{conditional variance}
$.  To ensure we have a Lebesgue density on $\mathbb{R}^{J-2}$, we work with a dimension-reduced coefficient vector by removing two coefficients and we denote this reduction with a bar over the vector and matrix. 

The final prior on the coefficients is given by a truncated normal prior distribution
$ \bar{\bm{\theta}}_d|\{\bm{A}\bm{\theta}_d = \bm{c}\} \sim \mathrm{TN}_{J-2}(\bar{\boldsymbol\xi}, \bar{\boldsymbol\Gamma}, \mathcal{T}), $
where $\bar{\bm{\theta}}_d$ is the dimension-reduced coefficient vector with the dimension-reduced mean vector $\bar{\boldsymbol\xi}$, dimension-reduced covariance matrix $\bar{\boldsymbol\Gamma}$, restriction 
$\mathcal{T} = \{\bar{\bm{\theta}}_d: \bar{\bm{F}}\bar{\bm{\theta}}_d + \bar{\bm{g}} >\mathbf{0}\}$.  Additionally, $\bar{\bm{F}}$ is the dimension-reduced matrix of the monotonicity constraints and $\bar{\bm{g}}$ is a dimension-reduced constant vector of the monotonocity constraints.  We denote the truncated normal distribution as  $\mathrm{TN}_p(\bm{\mu},\bm{\Sigma}, \mathcal{T})$ with mean $\bm{\mu}$, covariance matrix $\bm{\Sigma}$, restriction $\mathcal{T}$, and dimension $p$.  Any choice of $\boldsymbol\zeta$ is acceptable, but we use $\zeta_j = \nu + \tau \Phi^{-1}\big(\frac{j-0.375}{J-0.75+1}\big), \; j= 1, \ldots J$,  where $\nu $ is a constant, $\tau$ is a positive constant, and $\Phi^{-1}$ is the inverse of the cumulative distribution function of the standard normal distribution.  The idea is that by increasing the original components of the mean vector $\boldsymbol\zeta$, the truncation set $\mathcal{T}$ in the final prior of the B-spline coefficients will have a substantial prior probability.

Finally, we put an improper uniform prior on the mean $p(\bm{\mu})=\prod_{d=1}^p p_d(\mu_d)\propto 1$.   The resulting transformed variables, $\bm{Z}_d = \bm{Y}_d - \mu_d$, which are assumed to be distributed as $\N(\mathbf{0}, \bm{\Omega}^{-1})$ and $\bm{Y}_d =  \sum_{j=1}^J\bm{\theta}_{dj}B_j(\bm{X}_d)$, $d=1,\ldots,p$, are used to estimate the precision matrix and learn the structure of the underlying graph.

\subsection{Cholesky Decomposition Reformulated as Regression Problems}

We learn the structure of the precision matrix using a Cholesky decomposition.  Denote the Cholesky decomposition of $\boldsymbol\Omega$ as $\boldsymbol\Omega = \bm{L}\bm{L}'$, where $\bm{L}$ is a lower triangular matrix with elements $l_{kd}$.  Define the coefficients $\beta_{kd} = -l_{kd}/l_{dd}$ and the precision as $\phi_d = 1/\sigma_d^2 = l_{dd}^2$.  Then as described in \citep{wong_adaptive_2013}, the lower triangular entries of $\boldsymbol\Omega$, denoted as $\omega_{kd}$, are given by 
\begin{equation*}
\omega_{kd} = \sum_{m=1}^dl_{km}l_{dm} = \sum_{m=1}^d\beta_{km}\beta_{dm}\phi_m, \textup{ for } k \geq d.  
\end{equation*}

Accordingly, the multivariate Gaussian model $\bm{Z} \sim \N(\mathbf{0}, \boldsymbol\Sigma)$ is equivalent to the set of independent regression problems, 

$$\bm{Z}_d = \sum_{k>d}\beta_{kd}\bm{Z}_k + \epsilon_d, \; \epsilon_d \sim \N(0, \sigma_d^2), \; d = 1,\ldots, p,   $$
where $\beta_{kd}$ are the regression coefficients for $k = d+1, \ldots, p$ and $d = 1, \ldots, p$, and $\bm{Z}_d$ and $\bm{Z}_k$ are, respectively, the $d$th column and $k$th columns selected from matrix $\bm{Z}$.  We use the notation ${k>d}$ to indicate that the columns are  greater than the $d$th column.   

We use a standard conjugate noninformative prior on the variances.  We consider two different continuous shrinkage priors on the regression coefficients, the horseshoe prior and the Bernoulli-Gaussian prior.  Using these priors, we enforce a sparsity constraint along the rows of the lower triangular matrix.  The sparsity constraint is one in which the global sparsity parameter of the continuous shrinkage prior is scaled by $\sqrt{k}$, where $k > d$ and $d = 1,\ldots, p$.   Using this constraint, we expect that the precision matrix will be sparse through weak order invariance. The sparsity constraint is derived in the next subsection. 

\subsection{Sparsity Constraint}\label{SparsConstraint}

In order to ensure that the probability that an entry is nonzero (i.e. sparsity) remains roughly the same over different rows we cannot simply impose the same degree of sparsity on the rows of the Cholesky factor $\bm{L}$, but need to change it over rows appropriately. Denote the probability as $\textup{P}(\cdot)$.  To see how the Cholesky factor $\bm{L}$ depends on the row index, we observe that 
\begin{align*}
\textup{P}(\omega_{kd} \neq 0)
&= \textup{P}(\sum_m^p l_{km}l_{dm} \neq 0) \\ 
&= \textup{P}(l_{km}l_{dm} \neq 0 \textup{ for some } m) \\
&= 1- \textup{P}(l_{km}l_{dm} = 0 \textup{ for all } m) \\
&= 1-\textup{P}(\cap_{m=1}^{\min (k,d)}\{l_{km}l_{dm} = 0 \}) \\
&= 1- \prod_{m=1}^{\min (k,d)}\textup{P}(l_{km}l_{dm} = 0) \\
&= 1-\prod_{m=1}^{\min (k,d)} \{1 - \textup{P}(l_{km}l_{dm} \neq 0) \}\\
&= 1-\prod_{m=1}^{\min (k,d)} \{1 - \textup{P}(l_{km} \neq 0)\textup{P}(l_{dm} \neq 0) \} \\
&= 1-\{1 - \textup{P}(l_{km} \neq 0)\textup{P}(l_{dm} \neq 0)\}^{\min (k,d)}
\end{align*}

Let $\rho_k = $\textup{P}(Nonzero entry in the $k$th row of $\bm{L}$).  Then
$$\textup{P}(\omega_{kd}\ne 0)=1- (1-\rho_k\rho_d)^{\min(k,d)}.$$

If $k\sim d$, the expression is roughly $1-(1-\rho_k^2)^k$, which remains stable in $k$ if $\rho_k = c_p/\sqrt{k}$, where $c_p$ depends on $p$ but not on $k$.  Then we obtain the probability of non-zero to be $1-\exp(-c_p^2)$.  Further, choosing $c_p$ to be small for $p \rightarrow \infty$ makes the probability small, which is essential in higher dimension. 
We choose $\rho_k = $\textup{P}(nonzero in $k$th row)$ ={c}/({p\sqrt{k}})$, and tune the value of $c \in \{0.1,1,10\}  $ to cover a range of three orders of magnitude, i.e. $10^{-1}, \; 10^{0}, \; 10^{1} $.

\section{Variational Bayes Estimation}\label{VariationalBayesEstimation}

We observe $n$ independent samples, $\bm{X}_1, \ldots, \bm{X}_n$, from the nonparanormal model $\mathrm{NPN}(\bm{\mu},\bm{\Omega}^{-1},\bm{f})$ with a sparse $\bm{\Omega}$.  Based on these observations and the prior described in Section \ref{npn}, we intend to compute the posterior distribution to make inferences about $\bm{\Omega}$ and its structure, using the transformations $\bm{f}$.  Ideally, we would want to construct a complete variational Bayesian (VB) algorithm in which the B-spline coefficients, mean, and inverse covariance matrix are estimated all in one setting.  However, for our problem, there is no closed form solution for the truncated multivariate normal distribution, and closed form solutions are needed for the mean field variational Bayesian algorithms. Instead, we use an exact Hamiltonian Monte Carlo within Gibbs scheme to sample the B-spline coefficients and the mean.  We obtain the Bayes estimate of the B-spline coefficients, $\hat{\boldsymbol\theta}_d = \mathds{E}(\boldsymbol\theta_d|\bm{X}_1, \ldots, \bm{X}_n)$, and the Bayes estimate of the mean, $\hat{\mu}_d = \mathds{E}(\mu_d|\bm{X}_1, \ldots, \bm{X}_n)$, where $\mathds{E}(\cdot | \bm{X}_1, \ldots, \bm{X}_n)$ is the posterior mean operator.  We then apply the variational Bayesian method on the synthetic data obtained by transforming the original observations using the estimated transformations. Thus we estimate the transformed variables using
$$Z_{id} = \sum_{j=1}^J \hat{\theta}_{jd}B_j(X_{id}) - \hat{\mu}_d. $$ Ideally, instead of plugging in, one can obtain samples from the posterior distributions of the transformations and draw samples from the variational distributions of the precision matrix for each generated sample and accumulate them. However, even in moderately high dimension, such an approach is extremely computationally intensive. Since the posterior distributions of the transformations are consistent \citep{mulgrave_bayesian_2020}, they concentrate near the Bayes estimate. As the main goal is structure learning, the inability of the plug-in to assess the posterior variability of the transformations is not a highly deterring issue. Thus, although the proposed algorithm  is not fully Bayesian, it utilizes the strength of the variational Bayesian approach to identify conditional independence relations in a nonparanormal graphical model within a manageable time. While the variational inference generally underestimates the posterior variance \cite{blei_variational_2017}, since the goal of structure learning is to determine whether there are zeros or nonzeros in the precision matrix, we should still be able to determine whether the element is zero or not zero.  We illustrate the variational method on the Bernoulli-Gaussian prior, following the strategy described in \cite{ormerod_variational_2017}. Let the Bernoulli distribution be denoted as $\mathrm{Ber}$ and the inverse gamma distribution be denoted as $\mathrm{IG}(A,B)$ with shape parameter $A$ and scale parameter $B$.  We can describe the joint distribution by  
\begin{equation}
\label{SpikeSlabPrior}
\begin{aligned}
\bm{Z}_d | \boldsymbol\beta_{k>d}, \boldsymbol\sigma, \boldsymbol\Upsilon_{k>d} &\sim \N(\bm{Z}_{k>d}\boldsymbol\Upsilon_{k>d}\boldsymbol\beta_{k>d},\sigma_d^2\bm{I}), \;
\beta_{kd} \sim \N(0, g^2)\\
\qquad \upsilon_{kd} &\sim \mathrm{Ber}(\rho_{kd}^{*}), \; 
\sigma_d^2 \sim \mathrm{IG}(A,B)\\
\end{aligned}
\end{equation}
for $d = 1,\ldots,p$, where $\boldsymbol\beta_{k>d} = (\beta_{d+1},\ldots, \beta_{p})$ is the vector of regression coefficients, $\bm{Z}_{k>d}$ is the matrix of transformations,  and $\boldsymbol\Upsilon_{k>d}$ is a binary indicator matrix of 0s and 1s that is modeled by the Bernoulli distribution with elements $\upsilon_{kd}$.  The hyperparameters $g^2$, $A,$ and $B$, are fixed, and $\rho_{kd}^{*} \in [0,1]$ controls the sparsity.  This variant of the spike-and-slab prior indirectly models sparsity on the regression coefficients by putting a binary indicator on the regression coefficients in the likelihood, instead of directly modeling sparsity on the regression coefficients.  As such, if $\upsilon_{kd} = 0$ for the Bernoulli-Gaussian prior, then $\beta_{kd}|\upsilon_{kd} \sim \N(0, g^2)$, unlike in usual spike-and-slab priors in which $\beta_{kd}$ would be equal to exactly 0.  We select $\rho_{kd}^{*}$ using a tuning procedure that incorporates the sparsity constraint and is discussed in Subsection \ref{tuning}. 

The joint posterior distribution that we aim to compute is 
\begin{equation*}
\begin{aligned}
p(\boldsymbol\beta, \boldsymbol\Upsilon, \boldsymbol\sigma^2| \bm{Z}) \propto \prod_{i=1}^n\prod_{d=1}^{p-1}p(Z_{id}|\bm{Z}_{i,k>d} , \boldsymbol\beta_{k>d},\boldsymbol\Upsilon_{k>d}, \sigma_d^2) \\
\quad \times p(\boldsymbol\beta_{k>d}) \, p(\boldsymbol\Upsilon_{k>d}) \, p(\sigma_d^2) \, p(Z_{ip} |\sigma_p^2) \, p(\sigma_p^2).
\end{aligned}
\end{equation*}
By plugging in the estimated transformed variables, we use a variational Bayesian algorithm to compute the posterior distribution of the sparse precision matrix.  Mean field variational Bayesian inference involves minimizing the Kullback-Leibler divergence between the true posterior distribution and a factorized approximation of the posterior.  Let $\boldsymbol\kappa$ represent the set of parameters in the model and $\bm{Z}$ represent the matrix of estimated transformed variables.  Then $p(\boldsymbol\kappa|\bm{Z})$ is approximated by $q(\boldsymbol\kappa) = \prod_{k=1}^K q_k(\boldsymbol\kappa_k)$, where $(\boldsymbol\kappa_1,\ldots,\boldsymbol\kappa_K)$ is a partition of $\boldsymbol\kappa$.  The optimal $q_k$ densities  satisfy
$$q_k(\boldsymbol\kappa_k) \propto \exp[\mathds{E}_{\setminus q_k(\boldsymbol\kappa_k)}\{\log p(\bm{Z}, \boldsymbol\kappa) \} ],  $$
where $\mathds{E}_{\setminus q_k(\boldsymbol\kappa_k)}  $ is the expectation with respect to all densities except $q_k(\boldsymbol\kappa_k)$ \citep{bishop_pattern_2006}.  The variational lower bound (VLB) for the marginal likelihood for $\bm{Z}$ is then given by 
$$\textup{VLB(q)} = \mathds{E}_q[\log \{ {p(\bm{Z}, \boldsymbol\kappa)}/{q(\boldsymbol\kappa)} \} ],  $$
where $\mathds{E}_q$ is the expectation with respect to the density  $q_k(\boldsymbol\kappa_k)$. Using the coordinate ascent method, optimizing each $q_k$ while holding the others fixed will result in the algorithm converging to a local maximum of the lower bound.  

Following \citep{ormerod_variational_2017}, the choice of factorization that we use for the VB approximation is
\begin{equation*}
\label{factorization}
q(\boldsymbol\beta, \boldsymbol\upsilon, \boldsymbol\sigma^2) = q(\sigma_p^2)\prod_{d = 1}^{p-1} q(\boldsymbol\beta_d) q(\sigma_d^2) \prod_{k = d+1}^{p}q(\upsilon_{kd}),
\end{equation*}

with, for some choice of parameters,
\begin{equation*}
\label{optimalq}
\begin{aligned}
q^{*}(\boldsymbol\beta_d) &\sim \N(\boldsymbol\alpha_d, \boldsymbol\Sigma_d), \;
q^{*}(\sigma_d^2) \sim \mathrm{IG}(A + \frac{n}{2}, s_d), \\
q^{*}(\upsilon_{kd}) &\sim \mathrm{Ber}(w_{kd}).
\end{aligned}
\end{equation*}

The parameters are obtained by the VLB with respect to them by coordinate ascents, called variational updates, which we can derive as in \cite{ormerod_variational_2017}. Introduce the notations $\textup{expit}(x) = {\exp(x)}/\{1 + \exp(x)\}$, and $ \textup{logit}(x) = \log({x}/({1-x}))$, and let the symbol $\circ$ denote the Hadamard product between two matrices. Then we have 
\begin{equation*}
\label{updates}
\begin{split}
\boldsymbol\Sigma_d &= [\tau_d(\bm{Z}_{k>d}'\bm{Z}_{k>d}) \circ \boldsymbol\Omega_d + g^{-2}\bm{I}]^{-1}, \\
\boldsymbol\alpha_d &= \tau_d(\tau_d \bm{W}_d\bm{Z}_{k>d}'\bm{Z}_{k>d}\bm{W}_d + \bm{D}_d)^{-1}\bm{W}_d\bm{Z}_{k>d}'\bm{Z}_d,\\
s_d &= B + \frac{1}{2}[\left \Vert \bm{Z}_d \right \Vert^2 - 2\bm{Z}_d'\bm{Z}_{k>d}\bm{W}_d\boldsymbol\alpha_d\\ &+ \textup{tr}\{(\bm{Z}_{k>d}'\bm{Z}_{k>d} \circ \boldsymbol\Omega_d)(\boldsymbol\alpha_d\boldsymbol\alpha_d' + \boldsymbol\Sigma_d) \} ], \\
\eta_{kd} &= \textup{logit}(\rho_{kd}^{*}) - \frac{\tau_d}{2}(\alpha_{kd}^2 + \Sigma_{k, k}) \left \Vert \bm{Z}_k \right \Vert^2\\ &+ \tau_d[\alpha_{kd}\bm{Z}_k'\bm{Z}_d - \bm{Z}_k'\bm{Z}_{l>k}\bm{W}_{l>k}(\boldsymbol\alpha_{l>k}\alpha_{kd} + \boldsymbol\Sigma_{l>k,k})], \\
s_p &= B + \frac{1}{2}[\left \Vert \bm{Z}_p \right \Vert^2,\quad 
w_{kd}  = \textup{expit}(\eta_{kd}), \\ 
\tau_d &= \frac{2A + n}{2s_d}
\end{split}
\end{equation*}
for $l = k+1,\ldots, p$, and $k = d+1, \ldots, p$.  Note that we use the notation $l>k$ to indicate that the columns are greater than the $k$th column.  In addition, $\bm{W}_d = \textup{diag}(\bm{w}_{k>d})$  where $\bm{w}_{k>d} = (w_{d+1},\ldots, w_p)$,  $\boldsymbol\Omega_d = \bm{w}_d\bm{w}_d' + \bm{W}_d(\bm{I} - \bm{W}_d)$, and $\bm{D}_d = \tau_d(\bm{Z}_{k>d}'\bm{Z}_{k>d}) \circ \bm{W}_d \circ (\bm{I} - \bm{W}_d)  + g^{-2}\bm{I}$. 

Using these optimal $q_k$ densities, the VLB simplifies to 
\begin{equation}
\label{VBLowerBound}
\begin{split}
 \textup{VLB}(\bm{Z}; \boldsymbol\rho) =& -\frac{pn}{2}\log(2\pi) + pA\log(B)\\
 &- p\log \Gamma(A) - (A + \frac{n}{2})\log s_p\\
 &+ p\log \Gamma(A + \frac{n}{2}) \\
 &+   \sum_{d=1}^{p-1} \Big\{\frac{\#(k>d)}{2} - \frac{\#(k>d)}{2}\log(g^2)\\  
 &- (A + \frac{n}{2}) \log(s_d) + \frac{1}{2}\log \left \vert \boldsymbol\Sigma_d \right \vert \\ &-\frac{1}{2g^2}\textup{tr}(\boldsymbol\alpha_d\boldsymbol\alpha_d' + \boldsymbol\Sigma_d) 
 + \sum_{k = (d+1)}^p[w_{kd} \log(\frac{\rho_{kd}^{*}}{w_{kd}})\\ &+ (1-w_{kd})\log(\frac{1-\rho_{kd}^{*}}{1-w_{kd}}) ] \Big \}.
 \end{split}
\end{equation}

The variational Bayesian algorithm is detailed in the Appendix.

\subsection{Tuning Procedure}
\label{tuning}
For every $(p-1)$ regression problem, we choose the parameter $\rho_{kd}^{*}$ based on the tuning algorithm described in detail in Section 4 of \cite{ormerod_variational_2017}.  In this section, we describe the changes that we made to add the sparsity constraint.  We use the $\rho$ discussed in \cite{ormerod_variational_2017} and multiply that value with $\rho_k = c/(p\sqrt{k})$ to incorporate the sparsity constraint discussed in Section \ref{SparsConstraint}.  Thus, for the fixed $\rho$ that was discussed in \cite{ormerod_variational_2017}, for our work, that translates to $\rho_k^{*} = \textup{expit}(-0.5\sqrt{n})/(p\sqrt{k})$.  Note that since the dimension $d$ is not changing for $\rho_k^{*}$, we do not need to include $c$ for tuning.  For the fixed $\bm{w}$ that was discussed in \cite{ormerod_variational_2017}, for our work, that translates to the fixed $\bm{w}_{k>d}$, and we select $\rho_{kd}^{*}=  (\textup{expit}(\iota_j)c_j)/(p\sqrt{k})$, where $c_j$ is taken from an equally spaced grid of 50 points between 0.1 and 10, and $\iota_j$ varies over an equally spaced grid of 50 points between $-15$ and $5$.  We replace the $c$ with $c_j$
which leads a grid of 50 values of $c_j$ between $0.1$ and $10$
instead of the three values of $c \in \{0.1, 1,10\}$ that was discussed in Section \ref{SparsConstraint}.  The variational lower bound for the tuning procedure is only based on the preceding $(p-1)$ regressions and not the regression relations that involve $Z_p$ and $\sigma_p^2$.

\section{MCMC Estimation through Horseshoe Prior}\label{MCMCHorseshoe} 
\subsection{Horseshoe Prior}
We use the horseshoe prior described in \cite{neville_mean_2014}, to shrink the $\beta$ coefficients:
\begin{equation}
\label{HorseshoePrior}
\begin{aligned}
\textbf{Z}_d|(\textbf{Z}_{k>d}, \boldsymbol\beta_{k>d},\sigma_d^2) &\sim \N(\textbf{Z}_{k>d}\boldsymbol\beta_{k>d}, \sigma_d^2\bm{I}), \\
\beta_{kd} | (\lambda_d^2, b_{kd}, \sigma_d^2) &\stackrel{\mathrm{ind}}{\sim} \N(0, \frac{\sigma_d^2 b_{kd} c^2 \lambda_d^2}{p^2 k} ),\\
 \lambda_d^2|a_d &\sim \mathrm{IG}(\frac{1}{2}, \frac{1}{a_d}),  \quad
a_d \sim \mathrm{IG}(\frac{1}{2}, 1), \\
b_{kd}|h_{kd} &\stackrel{\mathrm{ind}}{\sim} \mathrm{IG}(\frac{1}{2}, \frac{1}{h_{kd}}),\quad
h_{kd} \sim \mathrm{IG}(\frac{1}{2},1), \\
\sigma_d^2 &\sim \mathrm{IG}(A,B),  
\end{aligned}
\end{equation}
for $d = 1,\ldots,p$,  where $\boldsymbol\beta_{k>d} = (\beta_{d+1},\ldots, \beta_{p})$, $\bm{Z}_{k>d}$ is the matrix of transformations, and $A$ and $B$ are fixed hyperparameters.

The global scale parameter $\lambda$ is roughly equivalent to the probability of a nonzero element  \citep{van_der_pas_horseshoe_2014}.  We enforce the sparsity constraint using, $(\lambda_d c)/(p\sqrt{k})$.  Thus, since we are working with the squared parameter, the factor in the variance term for $\beta_{kd}$ is $(\lambda^2 c^2)/(p^2 k)$, where $c \in \{0.1,1,10\}$.  

The joint posterior distribution, corresponding conditional posterior distributions, and the sampling algorithm are provided in the Appendix.  

\section{MCMC Estimation through Bernoulli-Gaussian Prior}\label{MCMCBG}
\subsection{Bernoulli-Gaussian Prior}
We use the same Bernoulli-Gaussian prior described in \eqref{SpikeSlabPrior}.
The joint posterior distribution, corresponding conditional posterior distributions, and the sampling algorithm are provided in the Appendix.

 \section{Thresholding}\label{Thresholding}
The thresholding procedure that we consider for the method using the horseshoe prior \eqref{HorseshoePrior} is based on a  0-1 loss function described in \citep{wang_bayesian_2012} for classification under absolutely continuous priors.  Although this procedure is heuristic, it seems to perform well in practice.  Other thresholding rules may be used, such as those based on posterior credible intervals \citep{khondker_bayesian_2013}, information criterion \citep{kuismin_use_2016}, clustering \citep{li_variable_2017}, posterior model probabilities \citep{banerjee_bayesian_2015,mohammadi_bayesian_2015}, and projection predictive selection \citep{williams_bayesian_2018},
but we chose to focus on the 0-1 loss procedure for this study.  

\subsection{0-1 Loss Procedure}\label{Loss}

We find the posterior partial correlation using the precision matrices from the Gibbs sampler of the horseshoe prior \eqref{HorseshoePrior} and the posterior partial correlation using the standard conjugate Wishart prior.  The posterior samples of the partial correlation using the precision matrices from the Gibbs sampler are defined as 
\begin{equation*}
e_{kd,m} = \frac{-\omega_{kd,m}}{\sqrt{\omega_{kd,m}\omega_{dd,m}}},
\end{equation*}
where $\omega_{kd,m}$ is a Markov chain Monte Carlo (MCMC) sample from the posterior distribution of $\bm{\Omega}_m$, where $m = 1, \ldots, M$, $M$ is the number of MCMC samples, and $k,d = 1,\ldots,p$.  The posterior partial correlation using the standard conjugate Wishart prior is found by starting with the latent observation, $\bm{Z}_m,$ which is obtained from the MCMC output.  We put a standard Wishart prior on the precision matrix, $\boldsymbol\Omega_m \sim \mathrm{W}_p(3, \bm{I})$ \cite{wang_bayesian_2012}, where $\bm{I}$ is the identity matrix.  Note that this Wishart prior does not assume sparsity, but $\bm{Z}$ is obtained from the MCMC output assuming sparsity of the precision matrix.
Through conjugacy, the posterior distribution is $\boldsymbol\Omega_m \sim \mathrm{W}_p(n+3, (\bm{I} + \bm{S}_m)^{-1})$, where $\bm{S}_m = \bm{Z}_m'\bm{Z}_m$.  We then calculate the mean of the posterior distribution, $\bm{H}_m = \mathds{E}(\boldsymbol\Omega_m|\bm{Z}_m) = (n+3)(\bm{I} + \bm{S}_m)^{-1}$.  Finally, we compute the posterior samples of partial correlation coefficients by conjugate Wishart prior as 
\begin{equation*}
j_{kd,m} = \frac{-h_{kd,m}}{\sqrt{h_{kd,m}h_{dd,m}}}, 
\end{equation*}
where $h_{kd,m}$ stands for the $(k,d)$th element of $\bm{H}_m$. 

We link these two posterior partial correlations for the 0-1 loss method.  We claim the event $\{\omega_{kd,m} \ne 0\}$ if and only if 
\begin{equation}
\label{lossprocedure}
\frac{e_{kd,m}}{j_{kd,m}} > 0.5
\end{equation}
for $k,d = 1,\ldots,p$ and $m = 1,\ldots, M$.  The idea is that we are comparing the regularized precision matrix from the horseshoe prior to the non-regularized precision matrix from the Wishart prior.  If the absolute value of the partial correlation coefficient from the regularized precision matrix is similar in size or larger than the absolute value of the partial correlation coefficient from the Wishart precision matrix, then there should be an edge in the edge matrix.  If the absolute value of the partial correlation coefficient from the regularized precision matrix is much smaller than the absolute value of the coefficient from the Wishart matrix, then there should not be an edge in the edge matrix.

\section{Choice of Prior Parameters}\label{ChoicePriorParams}
For the precision matrix being estimated with a horseshoe prior \eqref{HorseshoePrior}, we need to select the value of the parameter $c$ which controls the sparsity.  We solve a convex constrained optimization problem in order to use the Bayesian Information Criterion (BIC), as described in \cite{dahl_maximum_2005, dahl_covariance_2008}.  First, we find the Bayes estimate of the  inverse covariance matrix, $\hat{\boldsymbol\Omega} = \mathds{E}(\boldsymbol\Omega | \bm{Z})$.  We also find the average of the transformed variables, $\bar{\bm{Z}} = M^{-1}\sum_{m = 1}^M\bm{Z}_m$, where $\bm{Z}_m$, $m = 1, \ldots, M$, are obtained from the MCMC output.  Then, using the sum of squares matrix $\bm{S} =\bar{\bm{Z}}'\bar{\bm{Z}}$, we solve for $\hat{\boldsymbol\Omega}_{\mathrm{MLE}}$, the maximum likelihood estimate of the inverse covariance matrix, 
\begin{equation*}
\underset{\boldsymbol\Omega}{\text{minimize }} 
-n\log\det \boldsymbol\Omega + \textup{tr}(\boldsymbol\Omega \textbf{S}), \quad 
 \text{subject to }
\mathcal{C}(\hat{\boldsymbol\Omega}),
\end{equation*}
where $\mathcal{C}$ represents the constraint that all elements of  $\hat{\boldsymbol\Omega}$ at the locations of the zeros of the estimated edge matrix from the MCMC sampler are zero. The estimated edge matrix from the MCMC sampler will be described in more detail in Section \ref{SimulationResults}.  For computational simplicity, in the code, we represent this problem as an unconstrained optimization problem as described in \cite{dahl_maximum_2005, dahl_covariance_2008}.  

Lastly, we calculate 
$\textup{BIC} = -2\ell(\hat{\boldsymbol\Omega}_{\mathrm{MLE}}) + k\log n  $, 
where $k = \#\mathcal{C}(\hat{\boldsymbol\Omega})$, the sum of the number of diagonal elements and the number of edges in the estimated edge matrix, and $-\ell(\hat{\boldsymbol\Omega}_{\mathrm{MLE}}) = -n\log \det \hat{\boldsymbol\Omega}_{\mathrm{MLE}} + \textup{tr}(\hat{\boldsymbol\Omega}_{\mathrm{MLE}}\textbf{S})$. 
We select the $c$ that results in the smallest BIC.

\section{Simulation Results}\label{SimulationResults} 
We conduct a simulation study to assess the performance of the proposed methods using the horseshoe MCMC, indicated as Horseshoe, Bernoulli-Gaussian MCMC, indicated as Bernoulli-Gaussian, and variational Bayesian algorithm, indicated as Variational Bayes.  We choose not to include to the Bayesian nonparanormal graphical model described in \cite{mulgrave_bayesian_2020} because we want to maintain the Cholesky decomposition across all comparisons of the proposed methods.  We compare the structure learning results of our methods to the nonparanormal graphical model \citep{liu_nonparanormal:_2009} and to a Bayesian Gaussian copula graphical model \citep{mohammadi_bayesian_2017}, indicated as the Bayesian Copula, in which the rank likelihood is used to transform the random variables with a uniform prior on the graph, a G-Wishart prior on the inverse correlation matrix, and estimation is used with the birth-death MCMC \citep{mohammadi_bayesian_2015}.  These competing methods all utilize a transformation of the data to learn the graphical structure.

We assess the performance of these methods by calculating sensitivity, specificity, and the Matthews correlation coefficient (MCC).  We assess the effect of the transformation functions of our proposed methods by calculating the scaled $L_1$-loss.  These metrics are detailed in Subsection \ref{perf}.  In this section, we describe the data generation process used to conduct the simulation study.

 The random variables, $Y_1,\ldots,Y_p$, are simulated from a multivariate normal distribution such that $Y_{i1},\ldots,Y_{ip} \stackrel{\mathrm{i.i.d.}}{\sim}  \N(\boldsymbol\mu, \boldsymbol\Omega^{-1})$ for $i = 1, \ldots, n$.  The means  $\boldsymbol\mu$ are selected from an equally spaced grid between 0 and 2 with length $p$.  We consider nine different combinations of $n, p,$ and sparsity for $\boldsymbol\Omega$:

\begin{itemize}
\item $p=25$, $n=25$, sparsity = $10\%$ non-zero entries in the off-diagonals;
\item $p=50$, $n=100$, sparsity = $5\%$ non-zero entries in the off-diagonals;
\item $p=100$, $n=300$, sparsity = $2\%$ non-zero entries in the off-diagonals;
\item $p=25,\; n=25$, AR(2) model;
\item $p=50$, $n=100$, AR(2) model;
\item $p=100$, $n=300$, AR(2) model;
\item $p=25,\; n=25$, circle model;
\item $p=50$, $n=100$, circle model;
\item $p=100$, $n=300$, circle model,
\end{itemize}
 where the circle model and the AR(2) model are described by the relations 
\begin{itemize} 
\item Circle model: $\omega_{ii} = 2,\; \omega_{i, i-1} = \omega_{i-1,i} = 1$, and $\omega_{1,p}=\omega_{p,1} = 0.9$;
\item AR(2) model: $\omega_{i,i} = 1, \; \omega_{i, i-1} = \omega_{i-1,i} =  0.5$ and $\omega_{i, i-2} = \omega_{i-2,i} =  0.25$.

\end{itemize}

The percent sparsity levels for $\boldsymbol\Omega$ are computed using lower triangular matrices that have diagonal entries normally
distributed with $\mu_{\textup{diag}} = 1$ and $\sigma_{\textup{diag}} = 0.1$, and 
non-zero off-diagonal entries  normally distributed
with $\mu_{\setminus \textup{diag}} = 0$ and $\sigma_{\setminus \textup{diag}} = 1$, where $\setminus$ denotes the complement of the set.  

The observed variables $\bm{X} = (X_1, \ldots,X_p)$ are constructed from the simulated variables $Y_1, \ldots, Y_p$.  The functions used to construct the observed variables are three cumulative distribution functions (c.d.f.s): asymmetric Laplace, extreme value, and stable.  Any values of the parameters for the c.d.f.s could be chosen, but instead of selecting 25, 50, and 100 sets of parameters, we automatically choose the values of the parameters to be the maximum likelihood estimates with the {\tt mle} function in MATLAB.  The values of the parameters for each of the c.d.f.s are the maximum likelihood estimates for the parameters of the corresponding distributions (asymmetric Laplace, extreme value, and stable), using the variables $Y_1, \ldots, Y_p$. 

We follow the procedure in \cite{mulgrave_bayesian_2020} to estimate the transformation functions.  The hyperparameters for the normal prior are chosen to be $\nu = 1, \tau = 1,$ and $\omicron^2 = 1$.   To choose the number of basis functions, we use the Akaike Information Criterion as described in \cite{mulgrave_bayesian_2020}.  Samples from the truncated multivariate normal posterior distributions for the B-spline coefficients are obtained using the exact Hamiltonian Monte Carlo (exact HMC) algorithm \citep{pakman_exact_2014}.  
The initial coefficient values, $\theta_{dj, \textup{initial}}$, for the exact HMC algorithm are calculated using quadratic programming as described in \citep{mulgrave_bayesian_2020}.  After finding the initial coefficient values $\boldsymbol\theta_d$, we construct initial values for $Y_{d, \textup{initial}} =  \sum_{j=1}^J\theta_{dj, \textup{initial}}B_j(X_{d})$ using the observed variables.  These initial values  $\bm{Y}_{\textup{initial}}$ are used to find the initial values for $\boldsymbol\Sigma, \boldsymbol\mu$, and $\boldsymbol\Omega$ for the algorithm, where $\boldsymbol\Sigma_{\textup{initial}} = \textup{cov}(\bm{Y}_{\textup{initial}}), \boldsymbol\mu_{\textup{initial}} = \bar{\bm{Y}}_{\textup{initial}}$, where $\bar{\bm{Y}}_{\textup{initial}}$ is the average of $\bm{Y}_{\textup{initial}}$, and $\boldsymbol\Omega_{\textup{initial}} = \boldsymbol\Sigma_{\textup{initial}}^{-1}$. 

For the part of the simulation study in which we do not estimate the transformation functions, the initial values for the Horseshoe, Bernoulli-Gaussian, and Variational Bayes algorithms are constructed from the observed variables, $\bm{X}$, with $\boldsymbol\Sigma_{\textup{initial}} = \textup{cov}(\bm{X}), \; \boldsymbol\mu_{\textup{initial}} = \bar{\bm{X}}$, where $\bar{\bm{X}}$ is the average of $\bm{X}$, and $\boldsymbol\Omega_{\textup{initial}} = \boldsymbol\Sigma_{\textup{initial}}^{-1}$.  Afterwards, the mean $\boldsymbol\mu$ and the precision matrix $\boldsymbol\Omega$ are estimated using the algorithms as described in the previous sections.

The hyperparameter $g^2$ for the Bernoulli-Gaussian prior and the Variational Bayes algorithm is fixed at 10. The hyperparameters $A$ and $B$ for the inverse gamma distribution for the Bernoulli-Gaussian prior, the Variational Bayes algorithm, and the horseshoe prior, are fixed at $A = B = 0.01$.  The initial value, $\tau^0$, where $t=0$, for the Variational Bayes algorithm is chosen to be 1000.  The threshold $\epsilon$ for stopping the Variational Bayes algorithm is set to $\epsilon = 10^{-6}$.  For the Variational Bayes algorithm and the MCMC algorithm using the Bernoulli-Gaussian prior, the tuning procedure described in Subsection \ref{tuning} is used to find the hyperparameter for the Bernoulli distribution, $\rho_{kd}^{*}$.  Since the vector $\bm{w}_{k>d}$ from the tuning procedure consists of only 0 and 1 values, it is used as the initial indicator vector $\boldsymbol\upsilon_d$ for the MCMC algorithm  using the Bernoulli-Gaussian prior.  The data matrix that is used as input for the tuning procedure is $\bm{Z}_{\textup{initial}} = \bm{Y}_{\textup{initial}} - \boldsymbol\mu_{\textup{initial}}$, which was described in the previous paragraphs. 

For the MCMC algorithm for the horseshoe prior, we consider three values of $c$ that are a range of three orders of magnitude: $c \in \{0.1,1,10\}$.  The value of $c$ that yields the lowest BIC was selected for the final estimates of the precision matrix and edge matrix.  The 0-1 loss procedure described in Subsection \ref{Loss}  was used to threshold the precision matrices and construct the edge matrices.

For the simulation study, we run 100 replications for each of the nine combinations and assess structure learning for each replication.  We collect $10000$ MCMC samples for inference after discarding a burn-in of $5000$. We do not apply thinning.  The Bayesian copula method is implemented using the {\tt R} package, {\tt BDGraph} \citep{mohammadi_bdgraph_2019} using the option ``gcgm''.  Posterior graph selection is done using Bayesian model averaging, the default option in the {\tt BDGraph} package, in which it selects the graph with links for which their
estimated posterior probabilities are greater than 0.5.  The nonparanormal graphical model is implemented using the {\tt R} package {\tt huge}  \citep{zhao_huge_2015} using the option ``truncation''.  The graphical lasso method is selected for the graph estimation and the default screening method, lossless \citep{witten_new_2011,mazumder_exact_2012}, is used.  Three regularization selection methods are used to find the estimated precision matrix and select the graphical model: the Stability Approach for Regularization Selection (StARS) \citep{liu_stability_2010}, the modified Rotation Information Criterion (RIC) \citep{lysen_permuted_2009}, and the Extended Bayesian Information Criterion (EBIC) \citep{foygel_extended_2010}.  The default parameters in the {\tt huge} package are used for each selection method.  As in \cite{liu_nonparanormal:_2009}, the number of regularization parameters used is 50 and they are selected among an evenly spaced grid in the interval [0.16, 1.2].  

The code for the proposed Bayesian methods is written in {\tt MATLAB} and sparse representations of the matrices are used when appropriate.  For the Variational Bayes algorithm, when calculating $w_{kd}^{*} = \textup{expit}(\eta_{kd})$, it is set to 0 if  $\exp(\eta_{kd})$ is below $ 2^{-52}$, which is {\tt eps}, the floating-point relative accuracy  in MATLAB, while $w_{kd}^{*}$ is set to 1 if $\exp(\eta_{kd})$ is equal to {\tt infinity} in MATLAB for numerical stability.   Infinity results from operations that lead to results too large to represent as conventional floating-point values.  Similar adjustments are also applied for the Bernoulli-Gaussian MCMC. The code is given in the Appendix.

\subsection{Performance Assessment}\label{perf}

We compute the Bayes estimate of the precision matrix $\hat{\boldsymbol\Omega} = \mathds{E}(\boldsymbol\Omega|\bm{Z})$ by averaging all MCMC samples after burn-in, or the Variational Bayes estimate by averaging over 500 independent samples from the variational distribution.  The median probability model \citep{berger_optimal_2004} is used to obtain the Bayes  estimate of the edge matrix.  We find the estimated edge matrix by first using the 0-1 loss procedure to threshold the MCMC precision matrix samples, and then we take the mean of the thresholded precision matrices.  
If each off-diagonal element of the mean of the thresholded matrices is greater than 0.5, the element is registered as an edge in the estimated edge matrix, and if each off-diagonal element of the mean is not greater than 0.5, it is registered as no edge.  

We compute specificity (SP), sensitivity (SE), and Matthews Correlation Coefficient (MCC) to assess the performance of the graphical structure learning.  They are defined as follows:
\begin{align*}
\textup{Specificity} = \frac{\textup{TN}}{\textup{TN} + \textup{FP}}, \qquad \textup{Sensitivity} = \frac{\textup{TP}}{\textup{TP} + \textup{FN}},\\ 
\\
\textup{MCC} = \frac{\textup{TP} \times \textup{TN} - \textup{FP} \times \textup{FN}}{\sqrt{(\textup{TP} + \textup{FP})(\textup{TP} + \textup{FN})(\textup{TN} + \textup{FP})(\textup{TN} + \textup{FN})}},
\end{align*}
where TP is the number of true positives, TN is the number of true negatives, FP is the number of false positives, and FN is the number of false negatives. For all three metrics,  the higher the values are, the better is the classification. If there are models that are estimated to have no edges, they result in NaNs as MCC values.  

We also look at the effect of the transformation functions on parameter estimation for our methods.  We consider the scaled $L_1$-loss function, the average absolute distance, as a measure of parameter estimation.  
Scaled $L_1$-loss is defined as 
$$\textup{Scaled $L_1$-loss} = \frac{1}{p^2}\sum_k\sum_d \left \| \hat{\boldsymbol\Omega}_{kd} -\boldsymbol\Omega_{\textup{true},kd} \right \|
$$
where $\boldsymbol\Omega_{\textup{true},kd}$ stands for the true covariance matrix. Note that for the Bayesian Copula method, we use the estimated inverse correlation matrix and the true correlation matrix in place of the precision matrix for loss calculation.

We review the results of sensitivity, specificity, Matthews Correlation Coefficient (MCC), and
the scaled $L_1$-loss for each method using boxplots. In general, for sensitivity, specificity, and MCC,
the closer the boxplots are to one and the tighter the boxplots, the better the performance of the
method. For the scaled $L_1$-loss, the closer the boxplots are to zero and the tighter the boxplots, the
 better performance.

\begin{figure*}[!htbp]
    \centering
    \includegraphics[width=.9\linewidth]{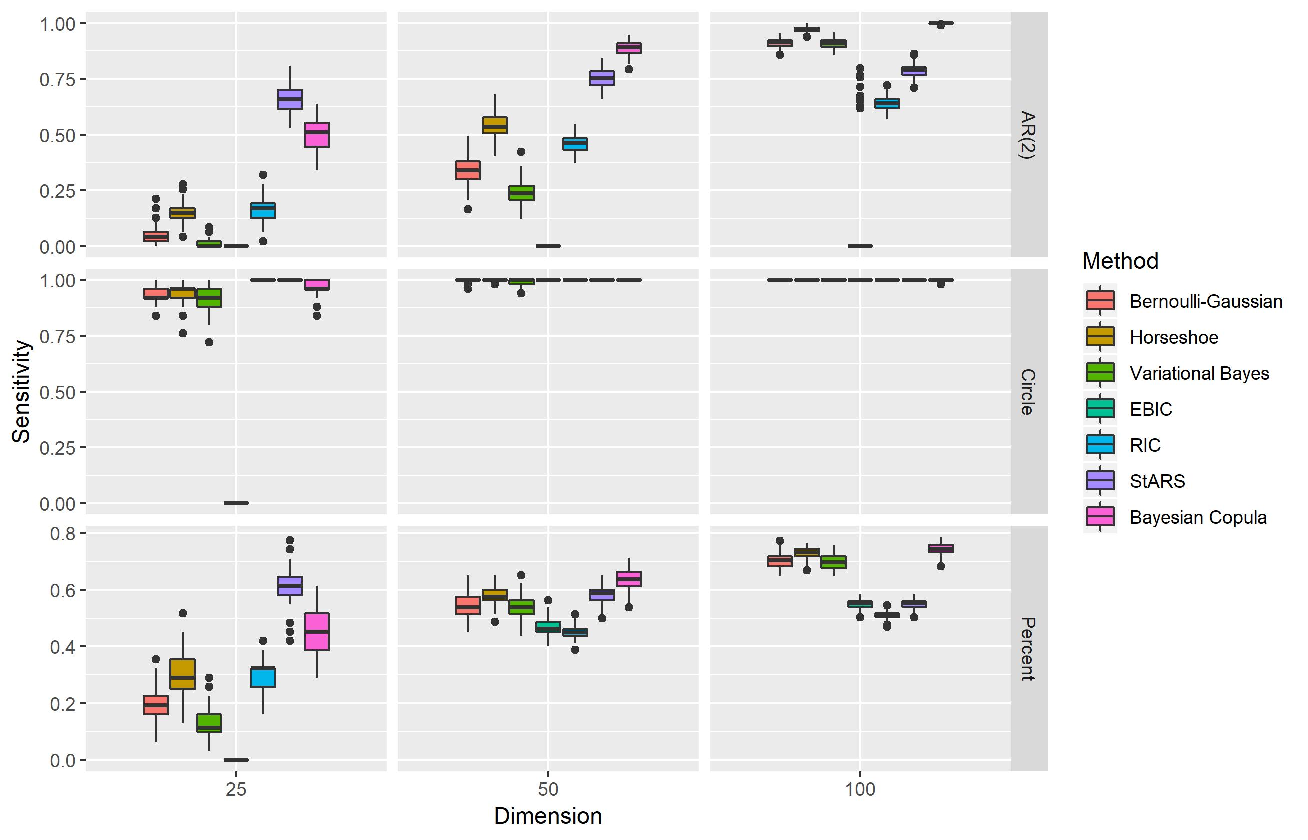}
    \caption{Boxplots of the sensitivity results for each of the methods for different structures of precision matrices. Percent refers to the 10\% model for dimension $p=25$, 5\% model for dimension $p=50$ and 2\% model for dimension $p=100$.}
   \label{fig:SensitivityResults}
   \vspace{1ex}
  \end{figure*}

First, we consider sensitivity. In Figure \ref{fig:SensitivityResults}\!for the $p = 25$ dimension and AR(2) model, the StARS model has the best sensitivity, followed with the Bayesian Copula model.  For $p=50$ and the AR(2) model, the Bayesian Copula performs the best, followed by the StARS model.  Notably, the proposed methods perform better at the $p=50$ dimension than at the $p=25$ dimension, with the Horseshoe method performing the third best.  Finally, for the $p=100$ dimension and AR(2) model, the Bayesian Copula method performs the best and the proposed methods perform second best, with the Horseshoe method performing the best and the Variational Bayes and Bernoulli-Gaussian methods performing third and fourth best.
  The Bayesian Copula is the best, the Horseshoe is the second best, and the Bernoulli-Gaussian and Variational Bayes methods are the third and fourth best, respectively.  For the $p=25$ dimension and the circle model, all methods are high-performing, but the RIC and StARS methods perform the best and the Bayesian Copula method is the third best.  For the $p=50$ and $p=100$ dimensions and the circle model, all methods perform similarly.  For the $p=25$ dimension and the 10\% model, the StARS method is the best and the Bayesian Copula method is the second best.  For the $p=50$ dimension and 5\% model, the Bayesian Copula method performs the best.  The Horseshoe and StARS methods perform similarly and are the second best, while the Variational Bayes and Bernoulli-Gaussian methods perform similarly and are the third best.  For the $p=100$ dimension and the 2\% model, the Bayesian Copula slightly outperforms the Horseshoe model, and the Bernoulli-Gaussian and Variational Bayes methods perform similarly at third best.

\begin{figure*}[!htbp]
    \centering
    \includegraphics[width=.9\linewidth]{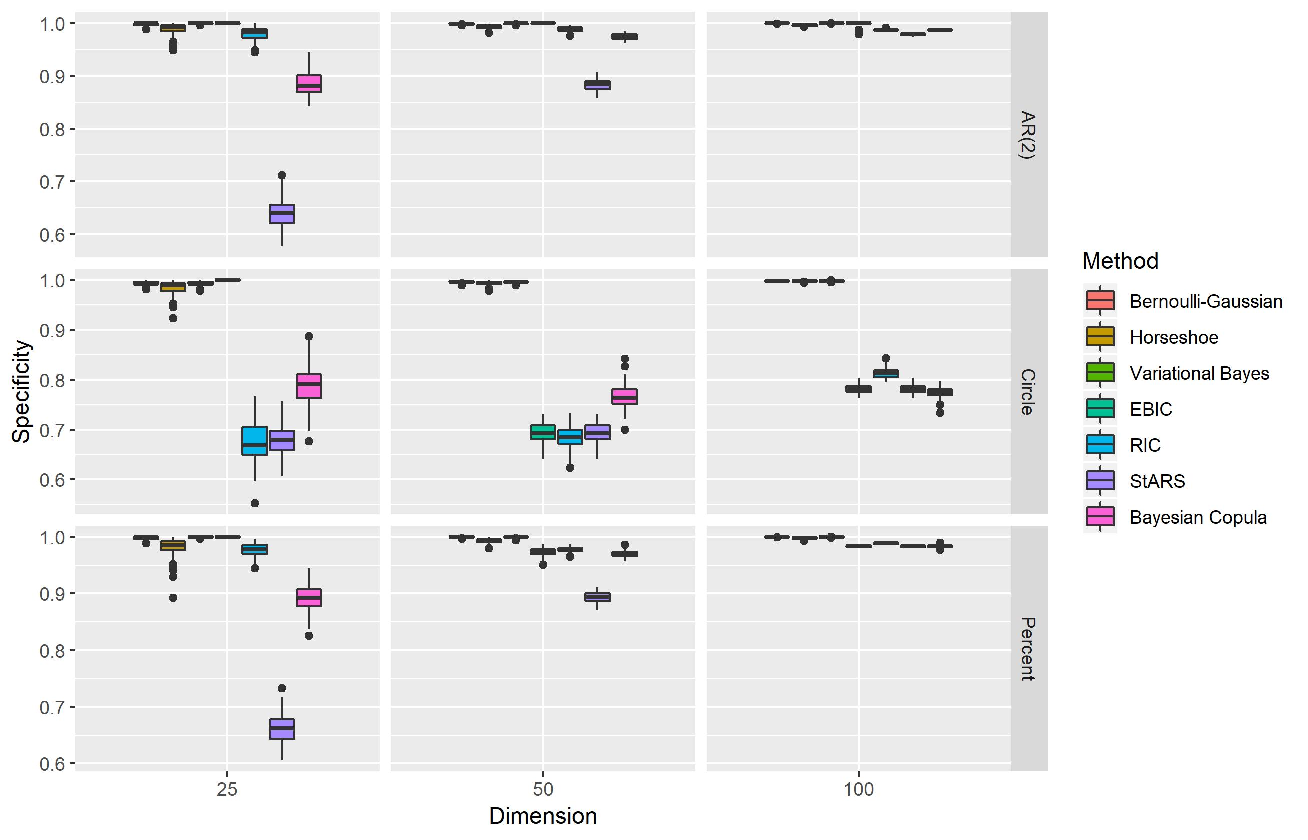}
    \caption{Boxplots of the specificity results for each of the methods for different structures of precision matrices. Percent refers to the 10\% model for dimension $p=25$, 5\% model for dimension $p=50$ and 2\% model for dimension $p=100$.}
    \label{fig:SpecificityResults}
    \vspace{1ex}
  \end{figure*}

Next, we review how the methods perform when considering specificity.  In Figure \ref{fig:SpecificityResults}\!for all dimensions and AR(2) model, the three proposed methods, Bernoulli-Gaussian, Horseshoe, and Variational Bayes methods, as well as the EBIC method, perform the best.  For the $p=25$ dimension and circle model, the three proposed methods, Bernoulli-Gaussian, Horseshoe, and Variational Bayes methods, as well as the EBIC method, perform the best.  For the $p=50$ and $p=100$ dimensions and the circle model,  the three proposed methods, Bernoulli-Gaussian, Horseshoe, and Variational Bayes methods, perform the best, outperforming all other methods.  For the $p=25$ dimension and the 10\% model, the three proposed methods, Bernoulli-Gaussian, Horseshoe, and Variational Bayes methods, as well as the EBIC method, perform the best.  For the $p=50$ dimension and 5\% model and the $p=100$ dimension and 2\% model,  the three proposed methods, Bernoulli-Gaussian, Horseshoe, and Variational Bayes methods, perform the best.

\begin{figure*}[!htbp]
    \centering
    \includegraphics[width=.9\linewidth]{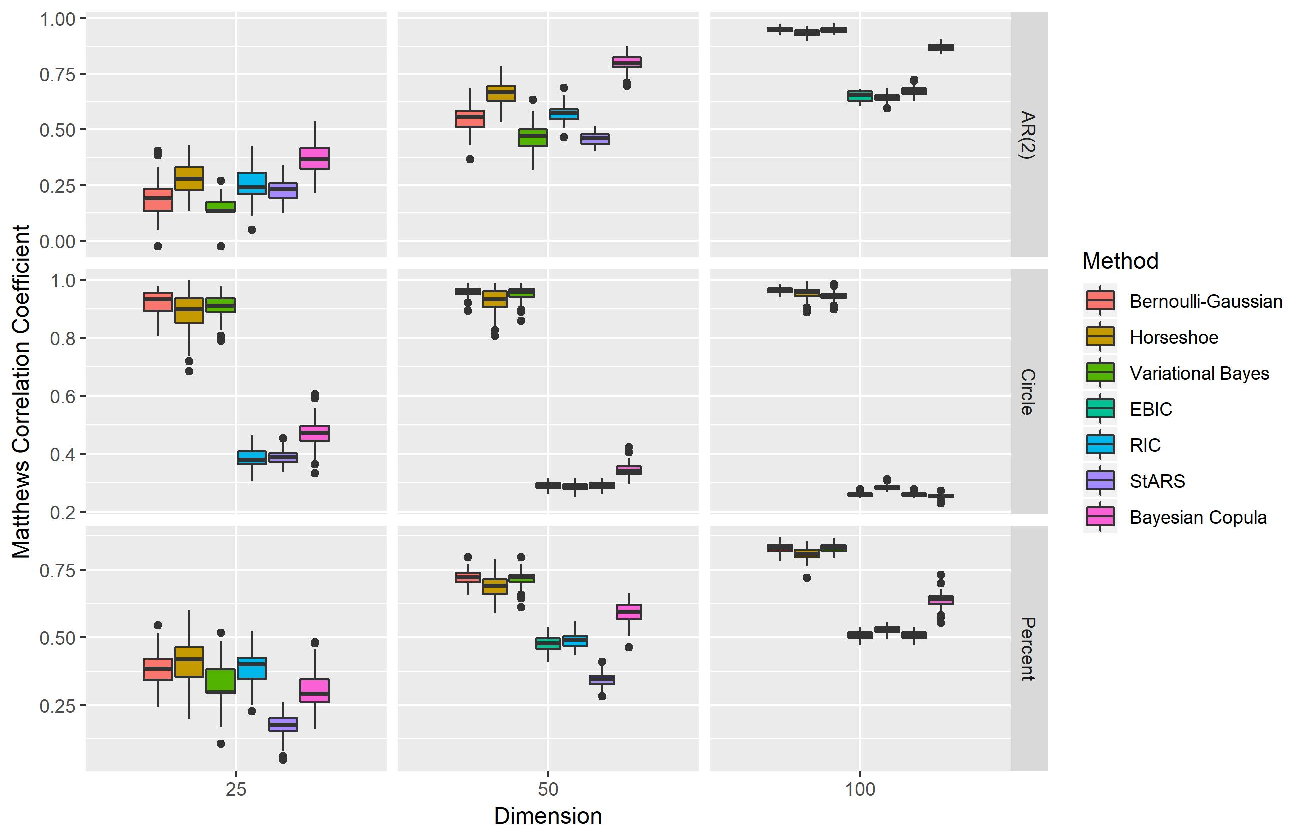}
    \caption{Boxplots of the Matthews correlation coefficient results for each of the methods for different structures of precision matrices. Percent refers to the 10\% model for the dimension $p=25$, 5\% model for the dimension $p=50$ and 2\% model for the dimension $p=100$.}
    \label{fig:MCCResults}
    \vspace{1ex}
  \end{figure*}

We consider the Matthews Correlation Coefficient to compare the overall performance of structure learning.  In Figure \ref{fig:MCCResults}\!for the $p=25$ and $p=50$ dimensions and the AR(2) model, the Bayesian Copula method performs the best and the Horseshoe method performs the second best.   No edges were selected by the nonparanormal model using EBIC for the sparsity models of dimension $p=25$ and for the $p=50$ AR(2) model.  For the $p=100$ dimension and the AR(2) model, the three proposed methods, Horseshoe, Bernoulli-Gaussian, and Variational Bayes methods, perform the best. For all dimensions of the circle model, the three proposed methods, Horseshoe, Bernoulli-Gaussian, and Variational Bayes methods, perform the best.  Lastly, for the $p=25$ dimension and 10\% model, the Horseshoe method performs the best, and the Bernoulli-Gaussian and RIC methods perform similarly and are the second best.  For the $p=50$ and 5\% model and $p=100$ and 2\% model, the three proposed methods, Horseshoe, Bernoulli-Gaussian, and Variational Bayes methods, perform the best. Thus, compared to competing methods, when considering overall structure learning, the proposed methods outperform the competing methods except in the cases of $p=25$ and $p=50$ and AR(2) model.  

\begin{figure*}[!htbp]
    \centering
    \includegraphics[width=.9\linewidth]{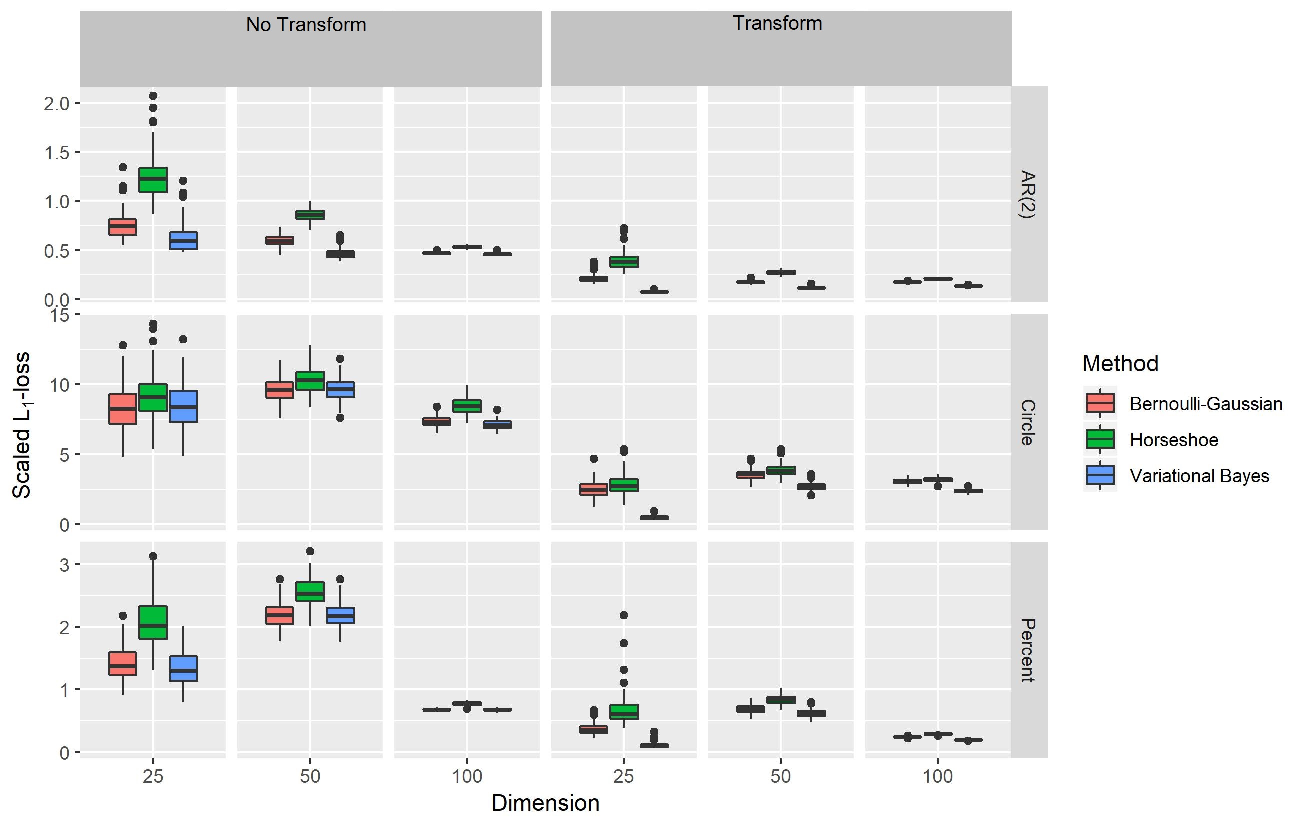}
    \caption{Boxplots of the scaled $L_1$-loss, with and without transformation for different structures of precision matrices. Percent refers to the 10\% model for the dimension $p=25$, 5\% model for the dimension $p=50$ and 2\% model for the dimension $p=100$.}
    \label{fig:ScaledLossResults}
    \vspace{1ex}
  \end{figure*}

Finally, in Figure \ref{fig:ScaledLossResults}\!we review the results of parameter estimation, using the scaled $L_1$-loss, for the three proposed methods.  We consider whether or not the transformation decreases the scaled $L_1$-loss.  For all three methods, the transformation functions resulted in a smaller scaled $L_1$-loss, implying  an improvement in parameter estimation.  Overall, the Horseshoe method had a higher scaled $L_1$-loss than the Bernoulli-Gaussian and Variational Bayes methods.  In addition, overall, the Variational Bayes method had a similar or lower scaled $L_1$-loss compared to the Bernoulli-Gaussian method.

Figures \ref{fig:SensitivityResults}\!to \ref{fig:ScaledLossResults}display the results.  The first three boxplots in the figures are the three proposed methods, Bernoulli-Gaussian, Horseshoe, and Variational Bayes, respectively.  Note that Percent refers to the 10\% model for dimension $p=25$, $5\%$ model for dimension $p=50$, and $2\%$ model for dimension $p=100$.


\section{Real Data Application}\label{RealData} 
For the real data application, we consider the data set based on the GeneChip (Affymetrix) microarrays for the plant \textit{Arabidopsis thaliana} originally referenced in  \citep{wille_sparse_2004}.  There are $n=118$ microarrays and $p=39$ genes from the isoprenoid pathway that are used.  For pre-processing, the expression levels for each gene, $x_i$ for $i = 1,\ldots, 118$, are log-transformed.  We study the associations among the genes using the Bayesian nonparanormal methods, the nonparanormal method of \cite{liu_nonparanormal:_2009}, and the Bayesian copula graphical model of \cite{mohammadi_bayesian_2017}.  These data are treated as multivariate Gaussian originally in \citep{wille_sparse_2004}.  

Using the same set-up as in the simulation study, we fit the Bayesian copula graphical model using the {\tt BDGraph} package and we fit the nonparanormal graphical model using the {\tt huge} package.  The {\tt BDGraph} package selected 211 edges using Bayesian model averaging.  The {\tt huge} package using the RIC selection resulted in 140 edges and using the StARS method resulted in 209 edges.  The EBIC-selected model results in no edges.

\begin{figure}[!htbp]
  \begin{subfigure}[b]{.75\linewidth}
    \centering
    \includegraphics[width=.7\linewidth]{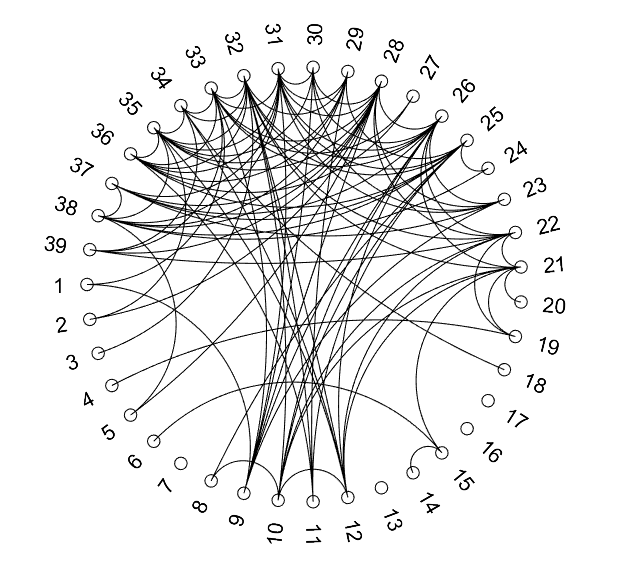}
    \caption{Variational Bayesian method.}
   \vspace{1ex}
  \end{subfigure}
  \begin{subfigure}[b]{.75\linewidth}
    \centering
    \includegraphics[width=.7\linewidth]{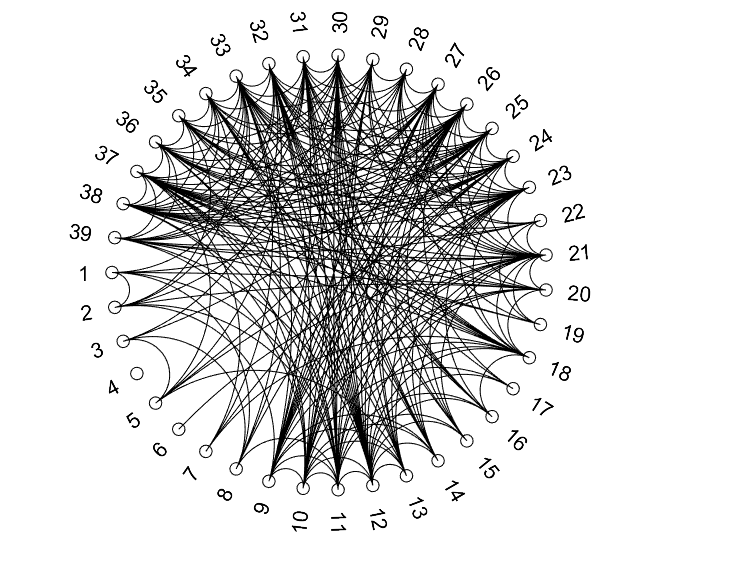}
    \caption{Horseshoe method.}
    \vspace{1ex}
  \end{subfigure}
    \begin{subfigure}[b]{.75\linewidth}
    \centering
    \includegraphics[width=.7\linewidth]{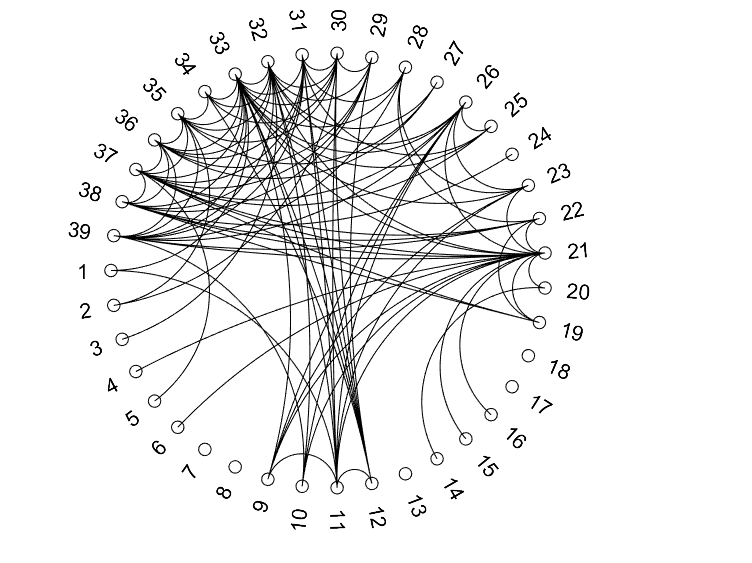}
    \caption{Bernoulli-Gaussian method.}
    \vspace{1ex}
  \end{subfigure}
    \caption{Comparison of selected graphs from the proposed methods using gene expression data.}
    \label{fig:RealDataProposedMethods}
\end{figure}

\begin{figure}[!htbp]
  \begin{subfigure}[b]{.75\linewidth}
    \centering
    \includegraphics[width=.7\linewidth]{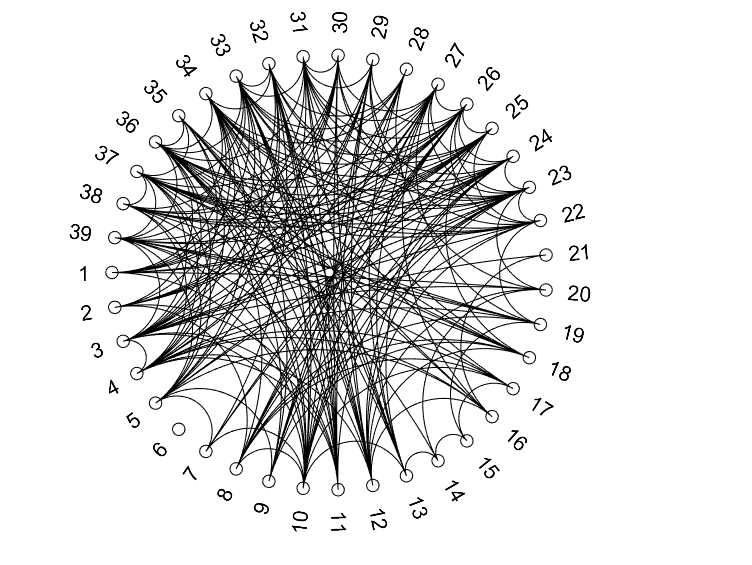}
    \caption{Bayesian copula method.}
   \vspace{1ex}
  \end{subfigure}
  \begin{subfigure}[b]{.75\linewidth}
    \centering
    \includegraphics[width=.7\linewidth]{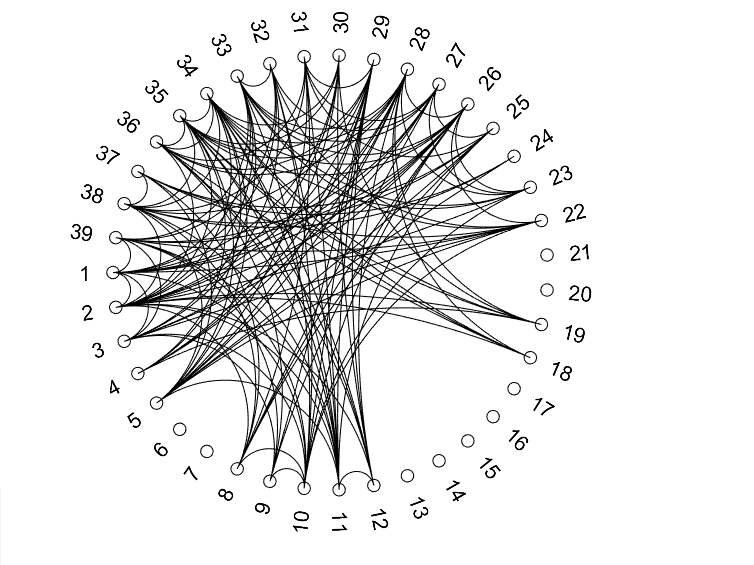}
    \caption{RIC method.}
    \vspace{1ex}
  \end{subfigure}
    \begin{subfigure}[b]{.75\linewidth}
    \centering
    \includegraphics[width=.7\linewidth]{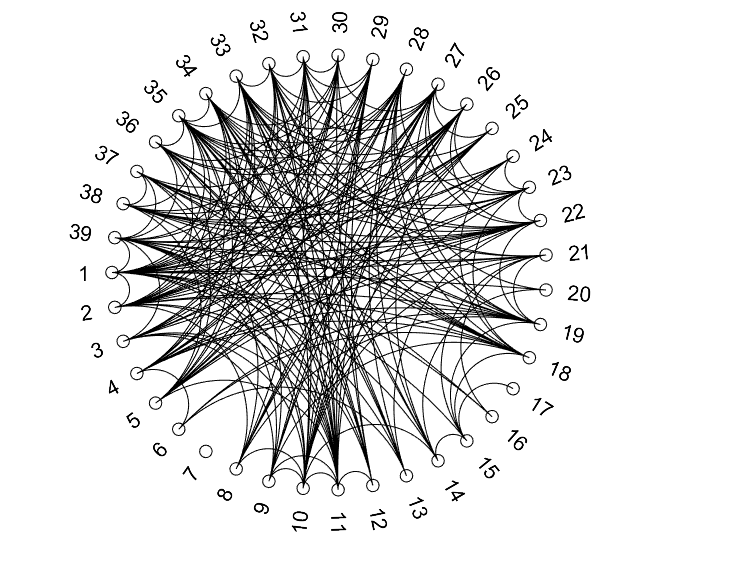}
    \caption{StARS method.}
    \vspace{1ex}
  \end{subfigure}
    \caption{Comparison of selected graphs from the existing methods using gene expression data.}
    \label{fig:RealDataExisingMethods}
\end{figure}

In order to construct the graphical models using our methods which use B-spline transformations, we converted the observations to be between 0 and 1 using the equation $({x - \min(x_i)})/({\max(x_i)-\min(x_i)})$.  The variational Bayes method results in 98 edges, the horseshoe prior based method results in 257 edges, and the Bernoulli-Gaussian prior based method results in 102 edges.  For $p=39$, convergence of the variational Bayes method can be achieved in about 26 minutes, the horseshoe prior based method in about 47 minutes for a given $c$, and the Bernoulli-Gaussian prior based method in about 52 minutes on a laptop computer with Windows operating system, 2.8 GHz of CPU, and 28 GB of RAM.  Figure \ref{fig:RealDataProposedMethods}shows the graphs of our proposed methods and Figure \ref{fig:RealDataExisingMethods}shows the graphs of the existing methods. 

Since we use a sparsity prior for each of the graphs, we consider the sparsity to compare the performance of the graphs.  The Variational Bayes and Bernoulli-Gaussian prior methods result in the sparsest graphs. The Horseshoe prior method results in the densest graph.  Out of the three proposed methods, the horseshoe prior method is the most sensitive method, so it appears for this data set, it is selecting more edges than the other models.  The Variational Bayes method is the fastest method out of the three proposed methods.  The Variational Bayes and Bernoulli-Gaussian prior methods proposed in this paper give sparser graphs than the Gaussian copula graphical model method, which uses a G-Wishart prior on the precision matrix.  Sparse graphs can aid in simpler scientific interpretation and could be used for further exploration, such as understanding the mechanisms involved in the isoprenoid pathway.

\section{Discussion}\label{Discussion}
We have introduced a Bayesian regression method to construct graphical models for continuous data that do not rely on a normality assumption.  The method assumes the nonparanormal structure, that under some unknown monotone transformations, the original observation vector reduces to a multivariate normal vector. The precision matrix of the transformed observations can be used to learn the graphical structure of conditional independence of the original observations. We use a prior distribution on the underlying transformations through a finite random series of B-splines with increasing coefficients that are given a multivariate truncated normal prior.  We incorporate the positive definiteness constraint on the precision matrix of the transformed variables by utilizing the Cholesky decomposition.  We consider two different priors based on the Cholesky decomposition, the Bernoulli-Gaussian prior and the horseshoe prior, and we impose a sparsity constraint. We use a variational Bayesian algorithm to learn the conditional independence relations more efficiently as well as use a traditional Gibbs sampling approach. The variational Bayesian approach and the approaches using Bernoulli-Gaussian and horseshoe priors result in most cases with better overall structure learning, measured using the Matthews correlation coefficient, than competing methods.  In addition, the variational Bayesian algorithm performs similarly to the proposed methods in terms of overall structure learning and parameter estimation.  Thus, it appears that information is not lost with the variational Bayesian algorithm and we have the potential to speed up the estimation of the Bayesian nonparanormal graphical model.  Lastly, when comparing the horseshoe to the Bernoulli-Gaussian methods, the horseshoe method has higher sensitivity and the Bernoulli-Gaussian methods perform similarly to the horseshoe in terms of specificity and overall structure learning and better in terms of parameter estimation.

Bayesian nonparanormal graphical models are flexible.  They can be used to estimate the elements of the precision matrix directly or via a Cholesky decomposition.  Researchers can try different sparsity priors on the precision matrix based on their interests and needs.  In addition, researchers can use a fully Bayesian approach to learn the graphical structure or employ a partially Bayesian approach to increase the speed in learning the structure without sacrificing much in quality.  The Bernoulli-Gaussian prior, used in the variational Bayesian method and the traditional Bayesian approach, resulted in the sparsest graphs using real data, which might be useful for researchers who would like greater variable reduction for data exploration.




\vskip 0.2in

\bibliography{Zotero.bib}


\section*{Acknowledgments}
We would like to acknowledge support for this project for the first author
from the National Science Foundation (NSF) Graduate Research Fellowship Program Grant No. DGE-1252376, the National Institutes of Health (NIH) training grant GM081057 and NSF grant DMS-1732842. Secondly, we would like to acknowledge partial support for this project for the second author by NSF grant DMS-1510238.

\subsection*{Financial disclosure}

None reported.

\subsection*{Conflict of interest}

The authors declare no potential conflict of interests.

\appendix

\section{Horseshoe Posterior}\label{AppendixHorseshoe}

The joint posterior distribution is,
\begin{equation*}
\label{HorseshoePosterior}
\begin{split}
p&(\boldsymbol\beta, \boldsymbol\lambda^2, \boldsymbol\sigma^2, \bm{a}, \bm{b}, \bm{h}, \boldsymbol\theta, \boldsymbol\mu| \bm{Z}) \propto  \prod_{i=1}^n\prod_{d=1}^{p-1}p(\sum_{j=1}^J\theta_{dj}B_j(X_{id})|\\
&\sum_{j=1}^J\theta_{k>d, j}B_j(X_{i,k>d}), \boldsymbol\beta_{k>d},\sigma_d^2) 
\times p(\boldsymbol\beta_{k>d}) \, 
p(\sigma_d^2)  \, p(\boldsymbol\theta_d) \, p(\mu_d)\\
&\times p(\bm{a}_d) \, p(\bm{b}_{k>d}) \, p(\bm{h}_{k>d})
 p(\lambda_d^2) \\
&\times p(\sum_{j=1}^J\theta_{pj}B_j(X_{ip})|\sigma_p^2) \, p(\sigma_p^2) \, p(\boldsymbol\theta_p) \, p(\mu_p). 
\end{split}
\end{equation*}

Then the corresponding conditional posterior distributions are given by 
\begin{equation*}
\begin{aligned}
\boldsymbol\beta_{k>d} \sim & \N[(\bm{Z}_{k>d}'\bm{Z}_{k>d} + \textup{diag}(\frac{p^2 k}{\lambda_d^2 \bm{b}_{k>d} c^2}))^{-1}\bm{Z}_{k>d}'\bm{Z}_d, \\ 
&\qquad \qquad \qquad \sigma_d^2(\bm{Z}_{k>d}'\bm{Z}_{k>d} + \textup{diag}(\frac{p^2 k}{\lambda_d^2 \bm{b}_{k>d} c^2}))^{-1}  ],\\
\lambda_d^2 \sim & \mathrm{IG}(\frac{\#(k>d)}{2} + \frac{1}{2}, \frac{1}{2}\boldsymbol\beta_{k>d}'\textup{diag}(\frac{p^2 k}{\sigma_d^2 \bm{b}_{k>d} c^2})\boldsymbol\beta_{k>d} + \frac{1}{a_d}), \\
a_d \sim & \mathrm{IG}(1, \frac{1}{\lambda_d^2} +1),\\
b_{kd} \sim & \mathrm{IG}(1, \frac{k \beta_{kd}^2 p^2}{2 \sigma_d^2\lambda_d^2 c^2} + \frac{1}{h_{kd}}), \\
h_{kd} \sim &\mathrm{IG}(1, \frac{1}{b_{kd}} + 1),\\
\sigma_d^2 \sim & \mathrm{IG}(\frac{n + \#(k>d)}{2} + A, \\
& \frac{1}{2}\left\Vert \bm{Z}_d - \bm{Z}_{k>d}\boldsymbol\beta_{k>d}\right \Vert^2 \\ 
&+ \frac{1}{2}\boldsymbol\beta_{k>d}'\textup{diag}(\frac{p^2 k}{\lambda_d^2 \bm{b}_{k>d} c^2 })\boldsymbol\beta_{k>d} + B),\\
\sigma_p^2 \sim & \mathrm{IG}(\frac{n}{2} + A, \frac{1}{2}\left \Vert \bm{Z}_p\right \Vert^2 + B).
\end{aligned}
\end{equation*}

Since sampling the $\boldsymbol\beta_{k>d}$ can be expensive for large $p$, we use an exact sampling algorithm for Gaussian priors based on data augmentation \citep{bhattacharya_fast_2016}.

\section{Bernoulli-Gaussian Posterior}\label{AppendixBG}
The joint posterior distribution is 
\begin{equation*}
\label{SpikeSlabPosterior}
\begin{split}
p&(\boldsymbol\beta, \boldsymbol\Upsilon| \bm{Z}) \propto  \prod_{i=1}^n\prod_{d=1}^{p-1}p(\sum_{j=1}^J\theta_{dj}B_j(X_{id})|\\
&\sum_{j=1}^J\theta_{k>d, j}B_j(X_{i,k>d}), \boldsymbol\beta_{k>d}, \boldsymbol\Upsilon_{k>d}, \sigma_d^2) 
\times p(\boldsymbol\beta_{k>d})  p(\boldsymbol\Upsilon_{k>d}) \\ &p(\sigma_d^2) \, p(\boldsymbol\theta_d) \, p(\mu_d)\, p(\sum_{j=1}^J\theta_{pj}B_j(X_{ip})|\sigma_p^2) \, p(\sigma_p^2)\, p(\boldsymbol\theta_p) \, p(\mu_p). 
\end{split}
\end{equation*}

Then the corresponding conditional posterior distributions are given by 
\begin{equation*}
\begin{split}
\boldsymbol\beta_{k>d}|\cdot &\sim \N[(\boldsymbol\Upsilon_{k>d}\bm{Z}_{k>d}'\bm{Z}_{k>d}\boldsymbol\Upsilon_{k>d} + \frac{\sigma_d^2 }{g^2}\bm{I})^{-1}\boldsymbol\Upsilon_{k>d}\bm{Z}_{k>d}'\bm{Z}_d, \\& \qquad \qquad \qquad (\boldsymbol\Upsilon_{k>d}\bm{Z}_{k>d}'\bm{Z}_{k>d}\boldsymbol\Upsilon_{k>d} + \frac{\sigma_d^2 }{g^2}\bm{I})^{-1}],\\
\upsilon_k|\cdot &\sim \mathrm{Ber}[\textup{expit}\{\textup{logit}(\rho_{kd}^{*}) - \frac{1}{2\sigma_d^2}\left\Vert\bm{Z}_k\right\Vert^2\beta_k^2 \\
&+ \frac{1}{\sigma_d^2}\beta_k \bm{Z}_k'(\bm{Z}_d - \bm{Z}_{l>k}\boldsymbol\Upsilon_{l>k}\boldsymbol\beta_{l>k})\}],\\
\sigma_d^2 &\sim \mathrm{IG}(\frac{n}{2} + A,  \frac{1}{2}\left \Vert \bm{Z}_d -\bm{Z}_{k>d}\boldsymbol\Upsilon_{k>d}\boldsymbol\beta_{k>d}\right\Vert^2 + B),\\
p(\sigma_p^2) &\sim \mathrm{IG}(\frac{n}{2} + A,  \frac{1}{2}\left \Vert \bm{Z}_p\right \Vert^2 + B),
\end{split}
\end{equation*}
where $k = d+1, \ldots, p$, and $d = 1,\ldots, p-1$.

Again, to sample $\boldsymbol\beta_{k>d}$ we used an exact sampling algorithm for Gaussian priors that invokes data augmentation \citep{bhattacharya_fast_2016}.

\begin{algorithm}[!ht]
\caption{Variational Bayesian Algorithm}\label{alg:VB}
\begin{algorithmic}[1]
\item Gibbs Sampler: Estimate $\boldsymbol\theta$ and $\boldsymbol\mu$ 
\For{\texttt{$d = 1: p$}}
\begin{enumerate}[(a)]
\item  Sample 
$\bar{\boldsymbol\theta}_d|(\bar{\boldsymbol\Theta}_{-d}, \bm{Y}, \boldsymbol\mu, \boldsymbol\Omega) \sim \mathrm{TN}\big( \bm{\gamma}, \bm{\Psi}, \{\bar{\bm{F}}_d\bar{\boldsymbol\theta}_d + \bar{\bm{g}}_d > \mathbf{0} \})$,  where $\bm{\gamma}$  and $\bm{\Psi}$ are defined in Section 3.1 of \cite{mulgrave_bayesian_2020}.
\end{enumerate}
 \EndFor
 \item Repeat Step 2 until convergence.
\item Compute $\hat{\boldsymbol\theta}_d = \sum_{m=1}^M\boldsymbol\theta_{dm}$ and $\hat{\mu}_d= \sum_{m=1}^M\mu_{dm}  $, where $M$ is the number of Markov Chain Monte Carlo samples.
\item Compute $Z_{id} = \sum_{j=1}^J\hat{\theta}_{jd}B_j(X_{id}) - \hat{\mu}_d$.
\item Using $\bm{Z}$, tune $\rho_{kd}^{*}$
 and find the initial values for $\bm{w}_{k>d}$ using the tuning procedure described in Subsection \ref{tuning}.
 \item Coordinate Ascent Variational Inference: To compute $\boldsymbol\Omega$
\begin{enumerate}[(a)]
\item Initialize with $t = 1, \bm{Z}_d, \bm{Z}_{k>d}, g^2, A, B, \tau_0, \boldsymbol\rho_d^{*}, \bm{w}_{k>d}$ where $\bm{w}_{k>d}^{(1)} \in [0,1]^{\#(k>d)}$
\For{\texttt{$d = 1: (p-1)$}}
\begin{itemize}
\item $ \bm{W}_d^{(t)} = \textup{diag}(\bm{w}_{k>d}^{(t)})$
\item $\boldsymbol\Omega_d = \bm{w}_d^{(t)}\bm{w}_d^{(t)'} + \bm{W}_d^{(t)}(\bm{I} - \bm{W}_d^{(t)})$ 
\item $\boldsymbol\Sigma_d^{(t)} = [\tau_d^{(t-1)}(\bm{Z}_{k>d}'\bm{Z}_{k>d}) \circ \boldsymbol\Omega_d^{(t)} + g^{-2}\bm{I}]^{-1}$
\item $\boldsymbol\mu_d^{(t)} = \tau_d^{(t-1)} \boldsymbol\Sigma_d^{(t)} \bm{W}_d^{(t)}\bm{Z}_{k>d}'\bm{Z}_d$
\item $ s_d = B + \frac{1}{2}[\left \Vert \bm{Z}_d \right \Vert^2 - 2\bm{Z}_d'\bm{Z}_{k>d}\bm{W}_d^{(t)}\boldsymbol\mu_d^{(t)} + \textup{tr}\{(\bm{Z}_{k>d}'\bm{Z}_{k>d} \circ \boldsymbol\Omega_d^{(t)})(\boldsymbol\mu_d^{(t)}\boldsymbol\mu_d^{(t)'} + \boldsymbol\Sigma_d^{(t)}) \} ]$
\item $\tau_d^{(t)} = \frac{2A + n}{2s_d}$
\item $\bm{w}_d^{*} = \bm{w}_d^{(t)} $
\For{\texttt{$k = (d+1): p$}}
 \begin{itemize}
\item $\eta_{kd} = \textup{logit}(\rho_{kd}^{*}) - \frac{\tau_d^{(t)}}{2}((\mu_k^{(t)})^2 + \Sigma_{k, k}^{(t)}) \left \Vert \bm{Z}_k \right \Vert^2 + \tau_d^{(t)}[\mu_k^{(t)}\bm{Z}_k'\bm{Z}_d - \bm{Z}_k'\bm{Z}_l\bm{W}_l^{(t)}(\boldsymbol\mu_l^{(t)}\mu_k^{(t)} + \boldsymbol\Sigma_{l,k}^{(t)})]$
\item $w_{kd}^{*} = \textup{expit}(\eta_{kd})$
 \end{itemize}
\EndFor
\item $\bm{w}_d^{(t+1)} = \bm{w}_d^{*}$
\end{itemize}
 \EndFor
\item $s_p^{(t)} = B + \frac{1}{2}[\left \Vert \bm{Z}_p \right \Vert^2 $
\item Repeat (b)--(d) until $\left \vert \textup{VLB}(\bm{Z}, \boldsymbol\rho)^{(t)} - \textup{VLB}(\bm{Z}, \boldsymbol\rho)^{(t-1)} \right \vert < \epsilon.   $
\end{enumerate}
\item Sample $\boldsymbol\beta_d \sim \N(\boldsymbol\mu_d, \boldsymbol\Sigma_d) $, $\upsilon_{kd} \sim \mathrm{Ber}(w_{kd}) $, $\sigma_d \sim \mathrm{IG}(A + n/2, s_d)$, and $\sigma_p \sim \mathrm{IG}(A+ n/2, s_p)  $
\item Compute 
$l_{kd} = -\upsilon_{kd}\beta_{kd}/\sigma_d \textup{ and } l_{dd} = 1/\sigma_d. $
\item Compute
$\boldsymbol\Omega = \bm{L}\bm{L}'.  $
\end{algorithmic}
\end{algorithm}

\begin{algorithm}
\caption{Horseshoe Gibbs Algorithm}\label{alg:HS}
\begin{algorithmic}[1]
\item Gibbs Sampler: Estimate $\boldsymbol\theta$, $\boldsymbol\mu$, and $\boldsymbol\Omega$: 
\For{\texttt{d = 1,\ldots, p}}
\begin{enumerate}[(a)]
\item $\bar{\boldsymbol\theta}_d|(\bar{\boldsymbol\Theta}_{-d}, \bm{Y}, \boldsymbol\mu, \boldsymbol\Omega) \sim \mathrm{TN}\big( \bm{\gamma}, \bm{\Psi}, \{\bar{\bm{F}}_d\bar{\boldsymbol\theta}_d + \bar{\bm{g}}_d > \mathbf{0} \})$ 
where $\bm{\gamma}$  and $\bm{\Psi}$ are defined in Section 3.1 of \cite{mulgrave_bayesian_2020}.
\end{enumerate}
 \EndFor
\item Compute $Y_{id} = \sum_{j=1}^J\theta_{dj}B_j(X_{id}).$ 
\item  Sample $\boldsymbol\mu|(\bm{Y}, \boldsymbol\Omega) \sim \N_p(\bar{\bm{Y}}, \frac{1}{n}\boldsymbol\Omega^{-1}).$ 
\item Compute $Z_{id} = Y_{id} - \mu_d.$
\For{\texttt{d = 1,\ldots, p-1}}
\begin{enumerate}[(a)]
\item  Sample 
$\boldsymbol\beta_{k>d}|\sigma_d, \bm{b}_{k>d}, \lambda_d^2 \sim \N(\textbf{A}^{-1}\bm{Z}_{k>d}^T \bm{Z}_d, \sigma_d^2\textbf{A}^{-1}),$ where
$\textbf{A} = (\bm{Z}_{k>d}'\bm{Z}_{k>d} + \textup{diag}({p^2 k}/({\lambda_d^2 \bm{b}_{k>d} c^2}))):$

\begin{enumerate}[(i)]
\item Sample $t \sim \N(\mathbf{0}, \bm{D})$ and $\delta \sim \textup{Normal}(0, I_n)$, where $\bm{D} = \sigma_d^2\textup{diag}({\lambda_d^2\bm{b}_{k>d} c^2}/({p^2 k})).$
\item set $v = \boldsymbol\Phi t + \delta$, where $\boldsymbol\Phi = \textbf{Z}_{k>d}/\sigma_d.$
\item solve for $w$ in $ (\boldsymbol\Phi \bm{D}\boldsymbol\Phi' + I_n)w = (\alpha - v)$, where $\alpha =  \textbf{Z}_d/\sigma_d.$
\item set $\beta = t + \bm{D} \boldsymbol\Phi'w. $
\end{enumerate}

\item Sample $
\lambda_d^2 \sim \displaystyle\mathrm{IG}(\frac{\#(k>d)}{2} + \frac{1}{2}, \frac{1}{2}\boldsymbol\beta_{k>d}'\textup{diag}(\frac{p^2 k}{\sigma_d^2 \bm{b}_{k>d} c^2 })\boldsymbol\beta_{k>d} + \frac{1}{a_d}).$
\item Sample $a_d \sim \mathrm{IG}(1, \lambda_d^{-2} +1).$
\item Sample $b_{kd} \sim  \displaystyle \mathrm{IG}(1, \frac{p^2 k \beta_{kd}^2}{2 \sigma_d^2\lambda_d^2 c^2} + \frac{1}{h_{kd}}).$
\item Sample $h_{kd} \sim \mathrm{IG}(1, b_{kd}^{-1} + 1).$
\item Sample  $\sigma_d^2  \sim  \displaystyle\mathrm{IG}(\frac{n + \#(k>d)}{2} + A, 
\frac{1}{2}\left\Vert \bm{Z}_d - \bm{Z}_{k>d}\boldsymbol\beta_{k>d}\right \Vert^2 + \frac{1}{2}\boldsymbol\beta_{k>d}'\textup{diag}(\frac{p^2 k}{\lambda_d^2 \bm{b}_{k>d} c^2})\boldsymbol\beta_{k>d} + B).
$
\end{enumerate}
\EndFor
\item Sample 
\begin{equation}
\sigma_p^2 \sim \mathrm{IG}(\frac{n}{2} + A, \frac{1}{2}\left \Vert \bm{Z}_p\right \Vert^2 + B).
\end{equation}

\item Compute
$l_{kd} = -\beta_{kd}/\sigma_d \textup{ and } l_{dd} = 1/\sigma_d.$

\item Compute
$\boldsymbol\Omega = \textbf{L}\textbf{L}'.$
\item These steps are repeated until convergence.
\end{algorithmic}
\end{algorithm}

\begin{algorithm}
\caption{Bernoulli-Gaussian Gibbs Algorithm}\label{alg:BG}
\begin{algorithmic}[1]
\item Gibbs Sampler: Estimate $\boldsymbol\theta$, $\boldsymbol\mu$, and $\boldsymbol\Omega$: 
\For{\texttt{d = 1,\ldots, p}}
\begin{enumerate}[(a)]
\item  Sample 
$\bar{\boldsymbol\theta}_d|(\bar{\boldsymbol\Theta}_{-d}, \bm{Y}, \boldsymbol\mu, \boldsymbol\Omega) \sim \mathrm{TN}\big( \bm{\gamma}, \bm{\Psi}, 
\{\bar{\bm{F}}_d\bar{\boldsymbol\theta}_d + \bar{\bm{g}}_d > \mathbf{0} \}),$  where $\bm{\gamma}$  and $\bm{\Psi}$ are defined in Section 3.1 of \cite{mulgrave_bayesian_2020}.

\end{enumerate}
\EndFor
\item Compute $Y_{id} = \sum_{j=1}^J\theta_{dj}B_j(X_{id})$. 
\item  Sample $\boldsymbol\mu|(\bm{Y}, \boldsymbol\Omega) \sim \N_p(\bar{\bm{Y}}, \frac{1}{n}\boldsymbol\Omega^{-1})$. 
\item Compute $Z_{id} = Y_{id} - \mu_d$.
\For{\texttt{d = 1,\ldots, p-1}}
\begin{enumerate}[(a)]
\item  Sample 
$\boldsymbol\beta_{k>d}|\sigma_d, \boldsymbol\Upsilon_{k>d} \sim \N(\textbf{A}^{-1}\boldsymbol\Upsilon_{k>d}\textbf{Z}_{k>d}^T \textbf{Z}_d, \sigma_d^2\textbf{A}^{-1})$, where \;
$\textbf{A} = (\boldsymbol\Upsilon_{k>d}\bm{Z}_{k>d}'\bm{Z}_{k>d}\boldsymbol\Upsilon_{k>d} + \frac{\sigma_d^2 }{g^2}\bm{I}).  $

\begin{enumerate}[(i)]
\item Sample $t \sim \N(\mathbf{0}, \bm{D})$ and $\delta \sim \N(\mathbf{0}, \bm{I}_n)$, where $\bm{D} = g^2\bm{I}$;
\item set $v = \boldsymbol\Phi t + \delta$, where $\boldsymbol\Phi = \textbf{Z}_{k>d}\boldsymbol\Upsilon_{k>d}/\sigma_d$;
\item solve for $q$ in $ (\boldsymbol\Phi \bm{D} \boldsymbol\Phi' + \bm{I}_n)q = (\alpha - v)$, where $\alpha =  \textbf{Z}_d/\sigma_d$;
\item set $\beta = t + \bm{D}\boldsymbol\Phi'q$.
\end{enumerate}
\item Sample $\upsilon_k|\beta_k, \sigma_d \sim \mathrm{Ber}[\textup{expit}\{\textup{logit}(\rho_{kd}^{*}) - \frac{1}{2\sigma_d^2}\left\Vert\bm{Z}_k\right\Vert^2\beta_k^2 + \frac{1}{\sigma_d^2}\beta_k \bm{Z}_k'(\bm{Z}_d - \bm{Z}_{l>k}\boldsymbol\Upsilon_{l>k}\boldsymbol\beta_{l>k})\}].$
\item Sample $\sigma_d^2|\boldsymbol\beta_{k>d}, \boldsymbol\Upsilon_{k>d} \sim \mathrm{IG}( \frac{n}{2} + A, \frac{1}{2}\left \Vert \bm{Z}_d -\bm{Z}_{k>d}\boldsymbol\Upsilon_{k>d}\boldsymbol\beta_{k>d}\right\Vert^2 + B)$.
\end{enumerate}
\EndFor
\item Sample 
$\sigma_p^2|\bm{Z}_p \sim  \mathrm{IG}( \frac{n}{2} + A, \frac{1}{2}\left \Vert \bm{Z}_p\right \Vert^2 + B)$.

\item Compute
$l_{kd} = -\upsilon_{kd}\beta_{kd}/\sigma_d \textup{ and } l_{dd} = 1/\sigma_d.$

\item Compute
$\boldsymbol\Omega = \bm{L}\bm{L}'.  $
\item These steps are repeated until convergence.
\end{algorithmic}
\end{algorithm}

\section{GitHub Repository}
\label{GitHub}
\url{https://github.com/jnj2102/BayesianRegressionApproach}\\
The code used to run the methods described in this paper are available on GitHub.


\end{document}